\documentclass[twocolumn,aps,superscriptaddress]{revtex4}
\usepackage{amssymb}
\usepackage{amsmath}
\usepackage{graphicx}
\usepackage[normalem]{ulem}
\usepackage[dvips]{color}
\usepackage{appendix}
\usepackage{color}

\def\es0{$E_{\rm sym}(\rho_0)$}

\def\us0{$U_{\rm sym}(\rho_0,k_F)$~}

\def\l0{$L(\rho_0)$~}

\begin{document}

\title{Curvature-slope correlation of nuclear symmetry energy and its imprints on the crust-core transition, radius and tidal deformability of canonical neutron stars}

\author{Bao-An Li\footnote{Bao-An.Li@tamuc.edu}}
\affiliation{Department of Physics and Astronomy, Texas A$\&$M University-Commerce, Commerce, TX 75429, USA}
\author{Macon Magno\footnote{mmagno@leomail.tamuc.edu}}
\affiliation{Department of Physics and Astronomy, Texas A$\&$M University-Commerce, Commerce, TX 75429, USA}

\date{\today}

\begin{abstract}

\begin{description}
\item[Background:]
The nuclear symmetry energy $E_{sym}(\rho)$ encodes information about the energy necessary to make nuclear systems more neutron-rich. It is currently poorly known especially at supra-saturation densities but has broad impacts on properties of neutron stars, nuclear structures and reactions. While its slope parameter L at the saturation density $\rho_0$ of nuclear matter has been relatively well constrained by recent astrophysical observations and terrestrial nuclear experiments, its curvature $K_{\rm{sym}}$ characterizing the $E_{sym}(\rho)$ around $2\rho_0$ remains largely unconstrained. Over 520 calculations for $E_{sym}(\rho)$ using various nuclear theories and interactions in the literature have predicted several significantly different $K_{\rm{sym}}-L$ correlations. 

\item[Purpose:] If a unique $K_{\rm{sym}}-L$ correlation of $E_{sym}(\rho)$ can be firmly established, it will enable us to progressively constrain the high-density behavior of $E_{sym}(\rho)$ using the available and better constrained slope parameter L. We investigate if and by how much the different $K_{\rm{sym}}-L$ correlations may affect neutron star observables. We also examine if LIGO/VIRGO's observation of tidal deformability using gravitational waves from GW170817 and NICER's recent extraction of neutron star radius using high-precision X-rays can distinguish the different $K_{\rm{sym}}-L$ correlations predicted. 

\item[Method:]
A meta-model of nuclear Equation of States (EOSs) with three representative $K_{\rm{sym}}-L$ correlation functions is used to generate multiple EOSs for neutron stars. We then examine effects of the $K_{\rm{sym}}-L$ correlation on the crust-core transition density and pressure as well as the radius and tidal deformation of canonical neutron stars. 

\item[Results:]
We found that the $K_{\rm{sym}}-L$ correlation affects significantly both the crust-core transition density and pressure. It also has strong imprints on the radius and tidal deformability of canonical neutron stars especially at small L values. The available data from LIGO/VIRGO and NICER set some useful limits for the slope L but can not distinguish the three representative $K_{\rm{sym}}-L$ correlations considered.

\item[Conclusions:]
The $K_{\rm{sym}}-L$ correlation is important for understanding properties of neutron stars. More precise and preferably independent measurements of the radius and tidal deformability from multiple observables of neutron stars have the strong potential to help pin down the curvature-slope correlation, thus the high-density behavior of nuclear symmetry energy.
\end{description}
\end{abstract}

\maketitle

\section{Introduction}
The crust-core transition density and pressure in neutron stars (NSs) play significant roles in modeling NS observational properties \cite{Baym1971,Baym2}. In particular, they affect the fractional moment of inertia of a NS's crust closely related to the pulsar glitch phenomenon as well as the radius and quadrupole deformation of both isolated NSs and those involved in NS binary mergers. Regardless of the approaches used, determining the NS crust-core transition point involves both the first-order and second-order derivatives of the nucleon average energy in isospin-asymmetric nuclear matter with respect to the density of neutrons and protons. It thus has been known since the earlier 70's that the core-crust transition density and pressure are very sensitive to the fine details of the isospin and dense dependences of the nuclear Equation of State (EOS) \cite{Baym1971,Baym2,Arp72}. Indeed, extensive studies using various nuclear many-body theories and interactions have examined effects of various terms in the EOS and demonstrated the significant model dependence of the NS crust-core transition density and pressure, see, e.g., Ref. \cite{Li19} for a recent review. In particular, both the slope $L=3\rho_0[\partial E_{\rm{sym}}(\rho)/\partial\rho]|_{\rho=\rho_0}$ and curvature $K_{\rm{sym}}=9\rho_0^2[\partial^2 E_{\rm{sym}}(\rho)/\partial\rho^2]|_{\rho=\rho_0}$ of nuclear symmetry energy $E_{\rm{sym}}(\rho)$ at the saturation density $\rho_0$ of nuclear matter were found to play very important roles in determining the crust-core transition density and pressure. They both also affect significantly the radii and tidal deformations of canonical neutron stars \cite{Li19}. In fact, it is well known that NS radii are mostly determined by the nuclear pressure around $2\rho_0$ where the $K_{\rm{sym}}$ has the most important influence on the $E_{\rm{sym}}(\rho)$ and the corresponding pressure there \cite{Lattimer2000,Lattimer2001,Lat07,Li08}. In turn, imprints of $K_{\rm{sym}}$ on observational properties of NSs may help us further constrain the poorly known high-density behavior of nuclear symmetry energy. 

Unfortunately, it has been very challenging to constrain the density dependence of nuclear symmetry energy $E_{\rm{sym}}(\rho)$ especially at supra-saturation densities \cite{Tesym}. While some significant progresses have been made in experimentally constraining the slope parameter L characterizing mostly the $E_{\rm{sym}}(\rho)$ around $\rho_0$, the curvature $K_{\rm{sym}}$ dominating the behavior of $E_{\rm{sym}}(\rho)$ around $2\rho_0$ is much less constrained presently. More quantitatively,  a survey in 2013 of 28 previous analyses of terrestrial nuclear laboratory experiments and astrophysical observations found a fiducial value of $L=59\pm 16$ MeV \cite{BAL13}. It changed to $L = 58.7 \pm 28.1$~MeV in the 2016 survey of 53 more diverse analyses using more data \cite{Oer17}. While the uncertainty range increased the mean value of L remains about the same. These earlier analyses normally did not give any useful information about the $K_{\rm{sym}}$. On the other hand, extensive surveys of over 520 theoretical predictions available in the literature up to 2014 \cite{Dut12,Dutra2014} indicated that $K_{\rm{sym}}$ is in the range of $-400 \leq K_{\rm{sym}} \leq 100$ MeV \cite{Tews17,Zhang17}. While the latest Bayesian analyses in 2020 of both the LIGO/VIRGO tidal deformability data from observing GW170817 \cite{LIGO} and the NICER radius data from observing PSR J0030+0451 \cite{Riley19,Miller19} found the most probable values of $L=66_{-20}^{+12}$ and $K_{\mathrm{sym}}=-120_{-100}^{+80}$ at 68\% confidence level \cite{Xie19,Xie20a}. The extracted L is consistent with its fiducial value within the error bars. These analyses of the new astrophysical data clearly provided some useful constraints on the model predictions for $K_{\mathrm{sym}}$, but its uncertainty range remains quite large. 

Predictions of the nuclear EOS based on many-body theories subject to various constraints, such as empirical properties of nuclear matter at $\rho_0$, some properties of finite nuclei and/or terrestrial nuclear reactions, naturally introduce correlations among some features of the EOSs predicted. Indeed, strong correlations between $L$ and $K_{\rm sym}$ have been found in predictions by various nuclear energy density functionals and/or microscopic many-body theories, see, e.g., Refs. \cite{Dut12,Dutra2014,Tews17,Zhang17,Maz13,Lida14,Pro14,Col14,India17,Holt}. As discussed above, some reasonably tight experimental/observational constraints on the slope L already exist. Moreover, coming nuclear experiments as well as astrophysical observations will help narrow down further its uncertainties. Thus, if one can establish firmly the $K_{\rm{sym}}-L$ correlation, it can then help progressively constrain the $K_{\rm{sym}}$ using the available constraints on L. 

In this work, using the EOS meta-modeling approach of Refs. \cite{Zhang2018,Zhang19apj,Zhang19EPJA,Zhang20a} we study effects of the $K_{\rm{sym}}-L$ correlation on the crust-core transition density and pressure as well as its imprints on the radius and tidal deformability of canonical neutron stars of mass 1.4 M$_{\odot}$.  When necessary and possible, we also discuss if constraints on the NS tidal deformability from LIGO/VIRGO's observation of GW170817 \cite{LIGO} and constraints on the NS radius from NICER's recent observation of PSR J0030+0451 \cite{Miller19,Riley19} can help distinguish different $K_{\rm{sym}}-L$ correlations and/or how they may help constrain the L parameter. 

The rest of the paper is organized as follows. In the next section, we first outline the EOS meta-modeling method and three representative $K_{\rm{sym}}-L$ correlation functions available in the literature. We then examine effects of the $K_{\rm{sym}}-L$ correlation on the crust-core transition density and pressure by comparing results using the three different $K_{\rm{sym}}-L$ correlation functions in a large EOS parameter space allowed by all existing constraints. We then investigate imprints of the $K_{\rm{sym}}-L$ correlation on the radius and tidal deformability of neutron stars. A summary and conclusions are given at the end.

\section{Theoretical approach}\label{theory}
Here we outline the major components of our approach. We focus on the new features but skip most of the details that one can easily find in the literature. For completeness and ease of the following discussions, when necessary we also recall briefly some of the well established equations and methods we adopted here. 

\subsection{Parameterizing the EOS of neutron-rich nucleonic matter}
For neutron-rich nucleonic matter of neutron density $\rho_n$ and proton density $\rho_p$, it has an isospin asymmetry $\delta=(\rho_n-\rho_p)/\rho$ and density $\rho=\rho_n+\rho_p$. Its EOS can be written as \cite{Bom91}
\begin{equation}
E(\rho,\delta)=E_0(\rho)+E_{\rm{sym}}(\rho)\delta^{2} +\mathcal{O}(\delta^4)
\end{equation}
in terms of the energy per nucleon $E_0(\rho)\equiv E(\rho ,\delta=0)$ in symmetric nuclear matter (SNM) and the symmetry energy $E_{\rm{sym}}(\rho )$.
For a given EOS  $E(\rho,\delta)$ from a nuclear many-body theory, it is customary to Taylor expand both the $E_0(\rho)$ and $E_{\rm{sym}}(\rho)$
as functions of $(\rho-\rho_0)(3\rho_0)$ with coefficients given by their density derivatives at $\rho_0$. This approach is particularly useful in the traditional forward-modeling of various physics problems. Unfortunately, the coefficients predicted so far are still very model dependent and often show characteristically different correlations. 

In a different method that is almost opposite to the traditional approach mentioned above, independent of the nuclear many-body theories and interactions used and without knowing {\it a priori} the EOS, one can simply parameterize the SNM EOS $E_0(\rho)$ and the symmetry energy $E_{\rm{sym}}(\rho )$ as functions of $(\rho-\rho_0)/(3\rho_0)$ in the same form as if they are Taylor expansions of some known EOSs. By randomly generating the relevant parameters for the parameterized $E_0(\rho)$ and $E_{\rm{sym}}(\rho )$, one can mimic all available EOSs in the literature. By purposely 
parameterizing the $E_0(\rho)$ and $E_{\rm{sym}}(\rho )$ as if they are Taylor expansions, one can limit their parameter ranges to those of the Taylor coefficients predicted by all available nuclear many-body theories.
Such kinds of meta-modeling of nuclear EOSs \cite{MM1,MM2} or EOS generators \cite{Zhang2018,Zhang19apj,Zhang19EPJA,Zhang20a} have been found particularly useful in solving the NS inverse-structure problems. They have been used successfully in both the direct inversion of NS observables in the three-dimensional high-density EOS parameter space \cite{Zhang2018,Zhang19apj,Zhang19EPJA,Zhang20a} and the Bayesian inferences of EOS parameters from NS observables \cite{Xie19,Xie20a,France1,France2} or heavy-ion reaction data \cite{Xie20b}. In general, parameterized functions are necessary in all machine learning processes. It is advantageous to select functions such that one can make better use of available data/knowledge to set the prior ranges and probability distributions of the parameters. Our parameterizations of the $E_0(\rho)$ and $E_{\rm{sym}}(\rho )$ are based on this consideration.

We parameterize the $E_0(\rho)$ and $E_{\rm{sym}}(\rho )$ up to the third power of $(\rho-\rho_0)(3\rho_0)$ according to
\begin{eqnarray}\label{E0para}
E_{0}(\rho)&=&E_0(\rho_0)+\frac{K_0}{2}(\frac{\rho-\rho_0}{3\rho_0})^2+\frac{J_0}{6}(\frac{\rho-\rho_0}{3\rho_0})^3,\\
E_{\rm{sym}}(\rho)&=&E_{\rm{sym}}(\rho_0)+L(\frac{\rho-\rho_0}{3\rho_0})+\frac{K_{\rm{sym}}}{2}(\frac{\rho-\rho_0}{3\rho_0})^2\nonumber\\
&+&\frac{J_{\rm{sym}}}{6}(\frac{\rho-\rho_0}{3\rho_0})^3.\label{Esympara}
\end{eqnarray}
Values of $E_0(\rho_0)$ and $E_{\rm{sym}}(\rho_0)$ have no effect on the crust-core transition and have little effects on NS global properties. We thus fix them at their known empirical values of $E_0(\rho_0)=-15.9 \pm 0.4$ MeV \cite{Brown14} and $E_{\rm sym}(\rho_0)=31.6\pm 2.7$ MeV \cite{BAL13,Oer17}. 

In generating the EOSs, the parameters $K_0$, $J_0$, L, $K_{\rm sym}$ and $J_{\rm sym}$ 
are varied in ranges or fixed at specific values consistent with our currently knowledge from nuclear theories and experiments. For instance, the most probable incompressibility of symmetric nuclear matter is relatively well constrained to $K_0=240 \pm 20$ MeV \cite{shlomo06,Jorge10,Garg18}. We thus use three separate values of 220, 240 and 260 MeV for the $K_0$. The $J_0$ was predicted to be in the range of $-800 \leq J_{0}\leq 400$ MeV using various nuclear theories and forces \cite{Tews17,Zhang17}. Its most probable value was found to be $J_0=-165_{-45}^{+55}$ \cite{Xie20a} at 68\% confidence level from very recent Bayesian analyses of NS properties, while a value of $J_0=-215^{+20}_{-20}$ MeV \cite{Xie20b} was inferred from a recent Bayesian analysis of the collective flow and kaon production in relativistic heavy-ion collisions. Obviously, these Bayesian analyses have clearly narrowed down the predicted range of $J_0$, but still have relatively large errors. Fortunately, as it has been shown very recently in Ref. \cite{France2} in a Bayesian analysis using the EOS meta-modeling approach of Refs. \cite{MM1,MM2}, the crust-core transition density and pressure are insensitive to $J_0$ even when it is varied in an extremely large range between $-1000$ and $+1000$ MeV. Thus, in studying the crust-core transition we simply set $J_0=0$. While in constructing the core EOS for studying properties of NSs, we will fix it at a value sufficient to support NSs with a maximum mass of about 2 M$_{\odot}$, and be consistent with the results of the two Bayesian analyses mentioned above. 

Consistent with earlier findings, it was also shown in Ref. \cite{France2} that the crust-core transition density and pressure are very sensitive to the symmetry energy parameters 
by varying them independently within their large prior ranges of $10 \leq L \leq 80$ MeV, $-400 \leq K_{\rm{sym}} \leq 200$ MeV, and $-2000 \leq J_{\rm{sym}}\leq 2000$ MeV, respectively \cite{France2}. Some of these prior ranges are significantly wider than but consistent with those from surveying earlier analyses of both nuclear and astrophysical data \cite{BAL13,Oer17}. Interestingly, it was shown in Refs. \cite{France3,Fra-crust2} that the crust-core transition density is almost equally sensitive to the slope $L$ and curvature $K_{\rm{sym}}$ while the sensitivity to the $J_{\rm{sym}}$ parameter is weaker but appreciable. These useful prior knowledge and finings are considered when we vary the EOS parameters in our own study here.

\subsection{Slope-curvature correlation of nuclear symmetry energy}
\begin{figure}[htb]
\begin{center}
\vspace{0.7cm}
 \resizebox{0.48\textwidth}{!}{
 \includegraphics[width=\linewidth]{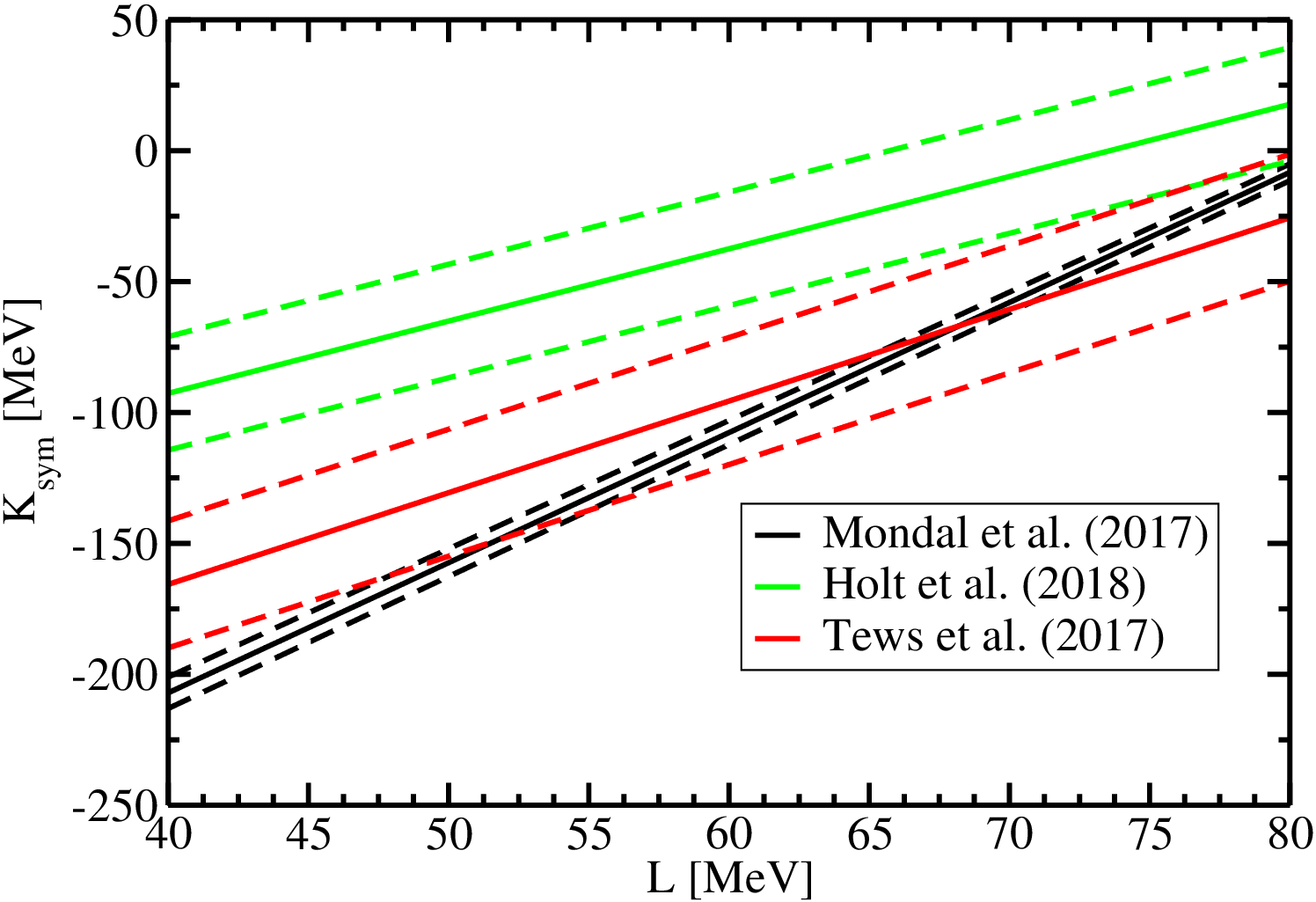}
 }
  \caption{(Color online) The $K_{\rm{sym}}-L$ correlation from Tews {\it et al.} \cite{Tews17} (red), Mondal {\it et al.} \cite{India17} (black) and Holt {\it et al.} \cite{Holt} (green), respectively. The solid lines are the means and the dashed lines are the upper and lower limits of each correlation.}\label{cor}
\end{center}
\end{figure}

Among the $K_{\rm{sym}}-L$ correlation functions found in the literature, the following one by Mondal {\it et al.} \cite{India17} 
\begin{equation}
   K_{\rm{sym}} = (-4.97 \pm 0.07)(3E_{\rm{sym}}(\rho_0)-L) + 66.80 \pm 2.14~ {\rm MeV}
\end{equation}
is based on probably the most number of theoretical predictions including 240 Skyrme Hartree-Fock (SHF) \cite{Dut12} and 263 Relativistic Mean-Field (RMF) calculations \cite{Dutra2014} compiled by Dutra {\it et al.}. 
Using the same inputs but restricting to predictions giving $0.149<\rho_0<0.17$ fm$^{-3}$, $-17 < E_0(\rho_0) < -15$ MeV, $25 < E_{\rm{sym}}(\rho_0) < 36$ MeV, and $180 < K_0 < 275$ MeV, Tews {\it et al.} \cite{Tews17} deduced the following correlation at 68\% confidence level
\begin{equation}
   K_{\rm{sym}} = 3.50 L -305.67 \pm 24.26~{\rm MeV}.
\end{equation}
Since the above two correlations stem from the same sets of model predictions albeit some additional selection criteria were used by Tews {\it et al.},  as shown in Fig. \ref{cor}, they largely overlap for $L>60$ MeV but show significant differences at lower L values. 

More recently, within the Fermi liquid theory with parameters benchmarked by chiral effective field theory predictions at sub-saturation densities, Holt and Lim \cite{Holt} derived 
the relations $L = 6.70 E_{\rm{sym}}(\rho_0)-148.60 \pm 4.37$ MeV and $ K_{\rm{sym}} = 18.50 E_{\rm{sym}}(\rho_0)-613.18 \pm 9.62$ MeV, leading to the 
$K_{\rm{sym}}-L$ correlation of 
\begin{equation}
   K_{\rm{sym}} = 2.76L - 203.07 \pm 21.69~{\rm MeV}.
\end{equation}
This correlation is shown as the green line in Fig. \ref{cor}. 

In the range of $40<L<80$ MeV consistent with its fiducial range from earlier surveys \cite{BAL13,Oer17}, 
the range of $K_{\rm{sym}}$ especially at lower L values in the Holt correlation is significantly higher than those in the other two cases. As we shall show, this has significant effects on properties of neutron stars.  
Holt {\it et al.} found that the largest source of uncertainty in their $K_{\rm{sym}}-L$ correlation is the assumed fiducial value of $K_0$ between 220 and 260 MeV.
We will thus also examine the role of $K_0$ in comparison with that of the $K_{\rm{sym}}-L$ correlation on the crust-core transition and radius of canonical neutron stars. We note that in this study we only used the means of the 
$K_{\rm{sym}}-L$ correlations shown in Fig. \ref{cor} without considering the uncertainty of each individual correlation.

\subsection{Thermodynamical method for finding the crust-core transition density in neutron stars}\label{kmu0}
Since the pioneering work of Baym {\it et al.} in 1971 \cite{Baym1971,Baym2}, the crust-core transition density and pressure in NSs have been studied extensively using several approaches starting from either 
the crust or core side. The most widely used one is by examining whether small density fluctuations will grow in the uniform core. This is often done by using the dynamical method 
considering the surface and Coulomb effects of clusters or its long-wavelength limit, i.e., the thermodynamical method, see, e.g., Refs. 
\cite{Pet84,Pet1,Kubis2007a,Kubis2007b,JXu1,JXu2,Dou00,Mou1,Cai12,Mou2,Sei14,Atta17,Rou16,Duc1a,Duc1b,Ava08,Gor10,Pea12,Sha15,Vid09,Cam10,Duc2,Bao1,Bao2,Gon17,Fang,Tsa,JP18}, 
or the RPA \cite{Horowitz2001,Car03,Fat10}. The crust-core transition has also been studied by comparing the free energy of clustered matter with that of the uniform matter either using various 
mass models within the Compressible Liquid Drop Mode \cite{Baym1971,Baym2,Pet84,Pet1,Dou00,Oyamatsu2007,Steiner08,Newton12} or the 
3D Hartree-Fock theory \cite{Newton09,Farrooh17} for nuclei on the Coulomb lattice using the Wigner-Seitz approximation.

Perhaps, the simplest approach is the thermodynamical method which we use here. In this approach \cite{Kubis2007a,Kubis2007b,Lat07}, the crust-core transition density is found by 
examining when the following effective incompressibility of the uniform NS core at $\beta$-equilibrium becomes negative 
(the corresponding speed of sound becomes imaginary), indicating the start of cluster formation (or the onset of spinodal decomposition)
\begin{eqnarray}\label{kmu1}
&&K_{\mu}= \rho^2 \frac{d^2 E_0}{d \rho^2} + 2 \rho \frac{dE_0}{d \rho}\\
&+& \delta^2
\left[ \rho^2 \frac{d^2 E_{\rm sym}}{d \rho^2}
+2 \rho \frac{d E_{\rm sym}}{d \rho} - 2 E^{-1}_{\rm sym}(\rho)
\left(\rho \frac{d E_{\rm sym}}{d \rho}\right)^2\right]. \nonumber
\end{eqnarray}
In terms of several EOS parameters, the crust-core transition density is determined by setting 
\begin{eqnarray}\label{kmu2}
&&K_{\mu}=\frac{1}{9} (\frac{\rho}{\rho_0})^2 K_0+ 2 \rho \frac{dE_0}{d \rho}\\&+& \delta^2
\left[\frac{1}{9} (\frac{\rho}{\rho_0})^2 K_{\rm{sym}}
+\frac{2}{3}\frac{\rho}{\rho_0}L
-2E^{-1}_{\rm sym}(\rho)(\frac{1}{3}\frac{\rho}{\rho_0}L)^2\right]=0. \nonumber
\end{eqnarray}
The last two terms in the bracket (isospin-dependent part) approximately cancel out, thus leaving the $K_{\rm{sym}}$ dominates \cite{Zhang2018}. Nevertheless, the $K_0$ and $L$ also play significant roles. 
Thus, different $K_{\rm{sym}}-L$ correlations are expected to affect the crust-core transition density. For earlier discussions on this issue, we refer the reader to Ref. \cite{Li19,Pro14,Newton12} and the references therein.  

It is interesting to note that Eq. (\ref{kmu2}) may have a second solution at a supra-saturation density besides the one at a sub-saturation density indicating the crust-core transition \cite{Kubis2007b}. This happens only in cases where the symmetry energy is super-soft (flat or decreasing with increasing density) at high densities when the $L$ is very small but the $K_{\rm{sym}}$ has a big negative value. In these cases, if the incompressibility $K_0$ of SNM is not high enough, then the negative contribution of the symmetry energy to the $K_{\mu}$ may cause the latter to decrease with increasing density. At some critical density, it will then reach zero again, indicating the onset of another dynamical instability. 
While by linking the liquid core EOS with a crust EOS at the crust-core transition density one has constructed a stable EOS up to the onset of the second instability, the physical meaning of the latter is currently not clear. Interestingly, it was speculated in Ref. \cite{Kubis2007b} that the second instability may indicate the start of another new phase, e.g., solidification, in the inner core of neutron stars. Without knowing how to model this essentially pure neutron matter core (as a result of the super-soft symmetry energy) at very high densities (above the generally expected hadron-quark transition density), we simply stop generating the EOS when the second instability happens (by default in the code by enforcing the dynamical stability condition in meta-modeling the EOS). This can happen before the causal limit or the dM/dR=0 point on the mass (M)-radius (R) curve is reached.

We found that the second instability happens mostly with the Mondal and Tews correlations giving large negative values of $K_{\rm sym}$ when L is around 40-50 MeV and $K_0$ is around 220-240 MeV. It does not happen with the Holt correlation as it gives significantly higher $K_{\rm sym}$ values. These can be easily understood from the competition of the different terms in Eq. (\ref{kmu2}). 
Thus, the three different $K_{\rm{sym}}$-L correlations with small L values affect significantly the maximum masses they predict. However, they have little effects on properties of canonical neutron stars as the second instability happens normally at very high densities reachable only in the core of massive neutron stars. For example, for the Mondal correlation with L=40 MeV,  $K_{\rm sym}$=-205 MeV, $J_{\rm sym}$=296.8 MeV, $K_0=220$ MeV and $J_0=-180$ MeV, the second instability happens at $6.82\rho_0$.  As a result, the maximum mass this EOS parameter set can support is about 1.9 M$_{\odot}$ while that with the Holt correlation is about 2.1 M$_{\odot}$ as we 
shall discuss in more detail in Section \ref{MR-r}. 

In our opinion, the possible appearance of the second instability when the symmetry energy is super-soft in the inner core of neutron stars is not a deficiency of the EOS meta-model we used. As pointed out in Ref. \cite{Kubis2007b}, there might be interesting new physics associated with the second instability. Without restrictions of the underlying energy density functionals in various nuclear many-body theories, the EOS meta-model can freely explore the entire EOS parameter space allowed by general physics principles. It can thus facilitate the study of previously unexplored areas of the EOS parameter space and the corresponding phases of neutron star matter.
Since the second dynamical instability may happen mostly at densities above the normally expected hadron-quark transition and we do not have a model for the EOS of the possible new phase above the second instability, we postpone the study on the possibly new physics associated with the latter to a future work. In this work, within the $npe\mu$ model enforcing the dynamical stability throughout neutron stars we focus on 
properties of canonical neutron stars that are not affected by the possible second instability.

\subsection{Constructing the EOS for neutron stars}

Within the $npe\mu$ model assuming NSs are made of neutrons, protons, electrons and muons at $\beta$-equilibrium under charge neutrality and dynamical stability conditions, the pressure 
\begin{equation}\label{pressure}
  P(\rho, \delta)=\rho^2\frac{d\epsilon(\rho,\delta)/\rho}{d\rho}
\end{equation}
is obtained from the energy density $\epsilon(\rho, \delta)=\epsilon_n(\rho, \delta)+\epsilon_l(\rho, \delta)$ with $\epsilon_n(\rho, \delta)$ and $\epsilon_l(\rho, \delta)$ being the energy densities of nucleons and leptons, respectively. 
While the $\epsilon_l(\rho, \delta)$ is calculated using the noninteracting Fermi gas model \citep{Oppenheimer39}, the $\epsilon_n(\rho, \delta)$ is from 
\begin{equation}\label{lepton-density}
  \epsilon_n(\rho, \delta)=\rho [E(\rho,\delta)+M_N]
\end{equation}
where $M_N$ is the average nucleon mass. 

\begin{figure*}[htb]
\begin{center}
%\vspace{-2cm}
 \resizebox{0.48\textwidth}{!}{
  \includegraphics[scale=0.4]{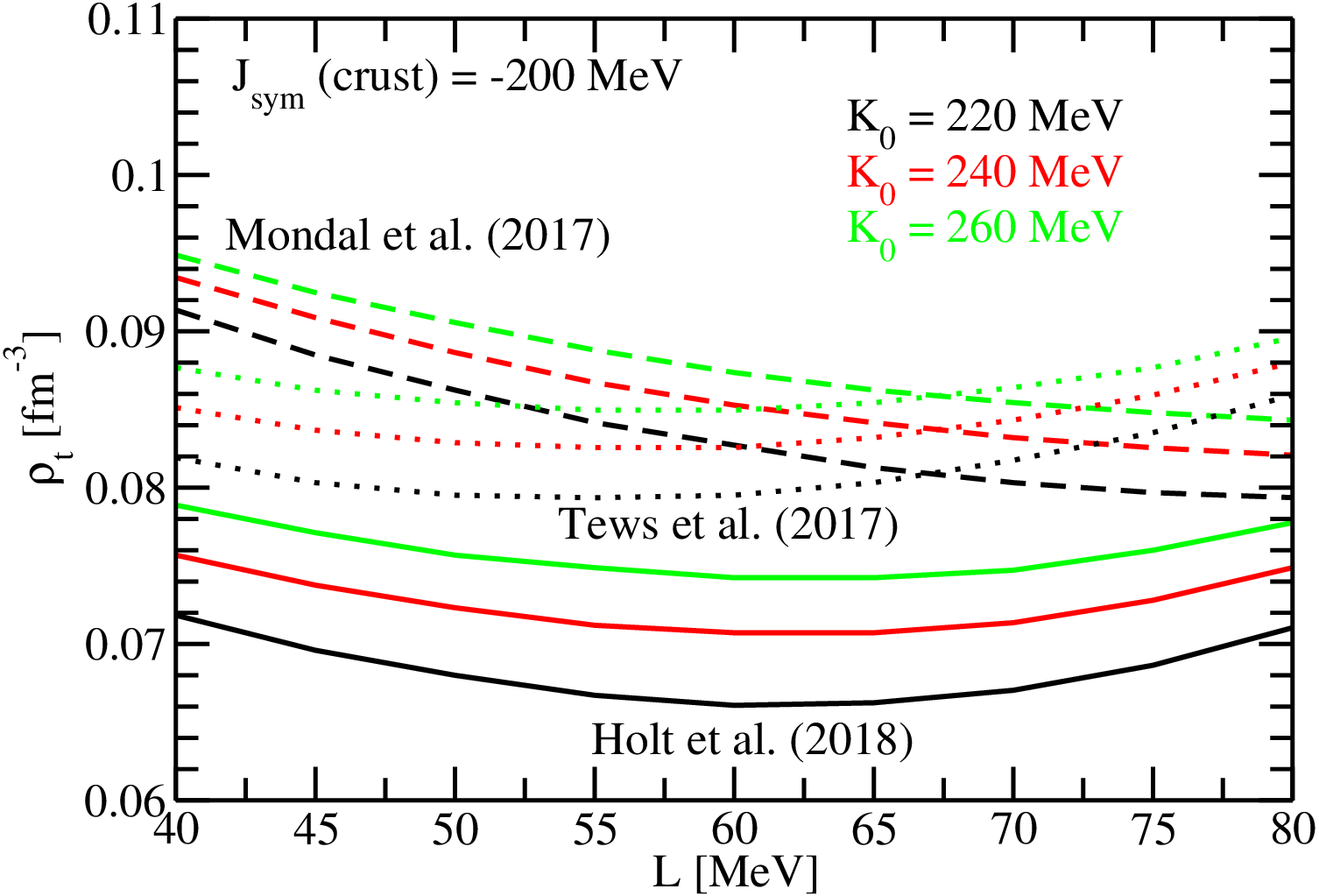}
 }
    \resizebox{0.48\textwidth}{!}{
  \includegraphics[scale=0.4]{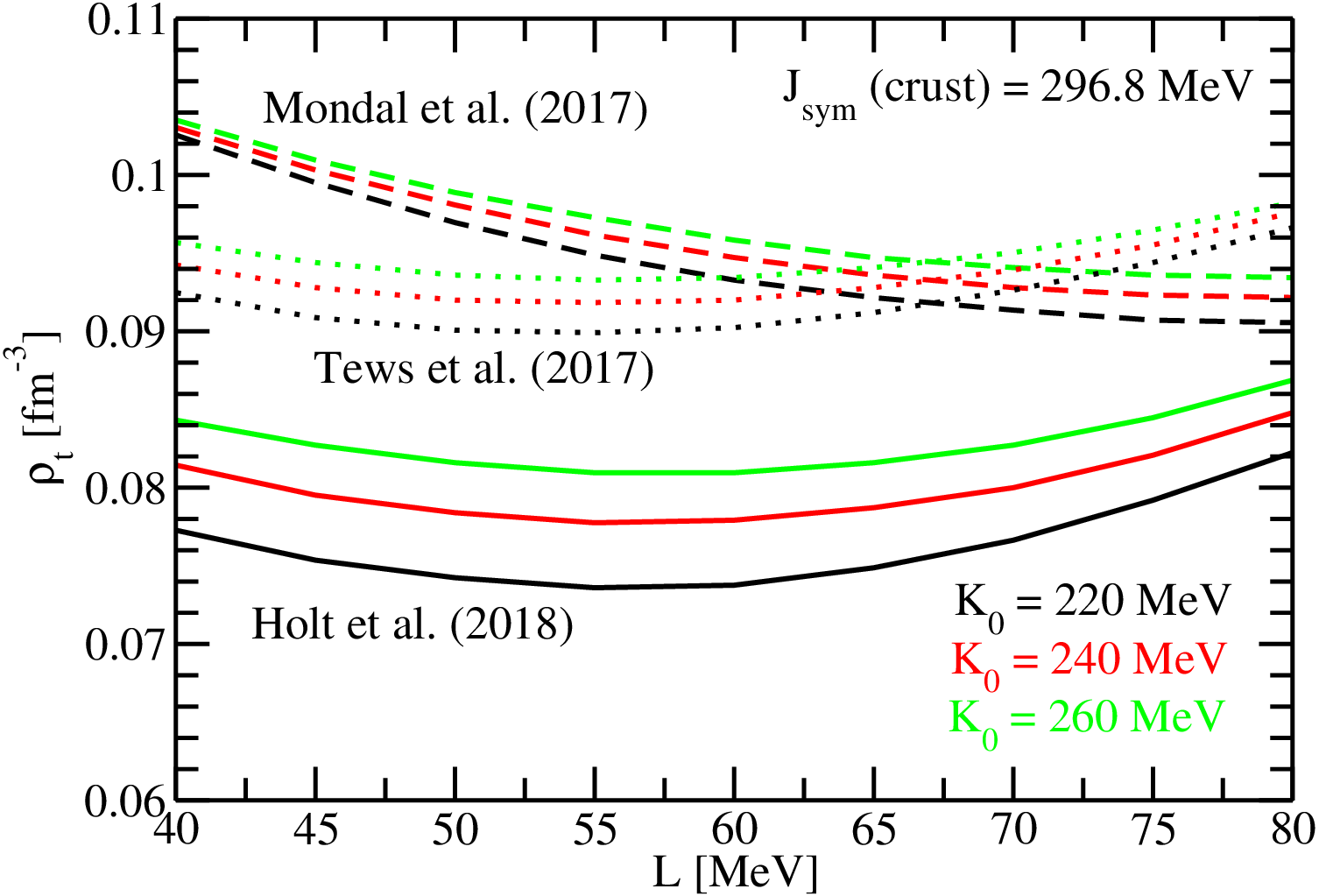}   }
  \caption{(Color online) The crust-core transition density as a function of $L$ with the indicated three different $K_{\rm{sym}}-L$ correlations with $K_0=220, 240$ and 260 MeV and $J_{\rm{sym}}(\rm{crust})=-200$ (left) and $+296.8$ MeV (right), respectively.}\label{densityL}
\end{center}
\end{figure*}

\begin{figure*}[htb]
\begin{center}
\vspace{0.5cm}
    \resizebox{0.48\textwidth}{!}{
 \includegraphics[scale=0.4]{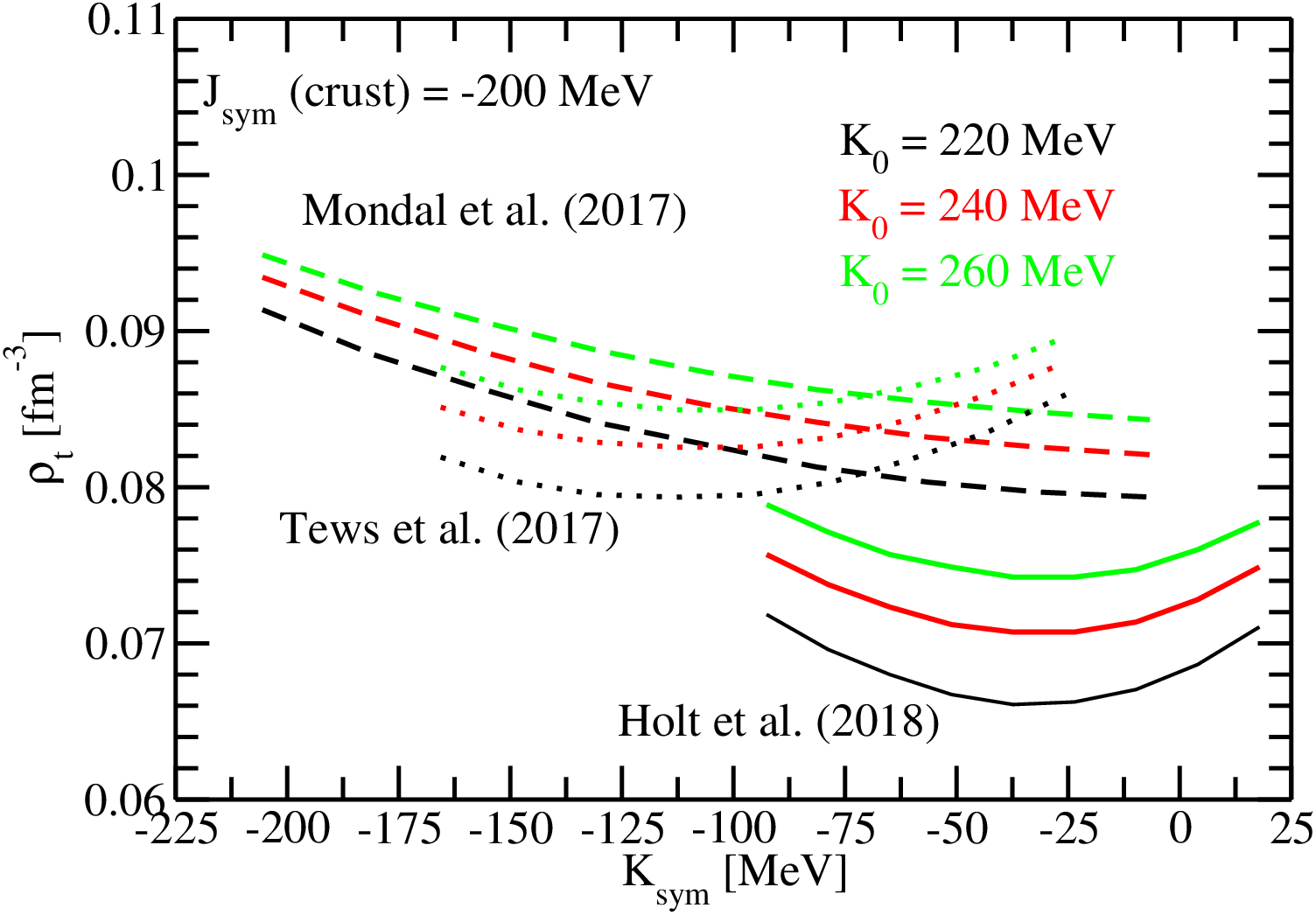}
 }
    \resizebox{0.48\textwidth}{!}{
  \includegraphics[scale=0.4]{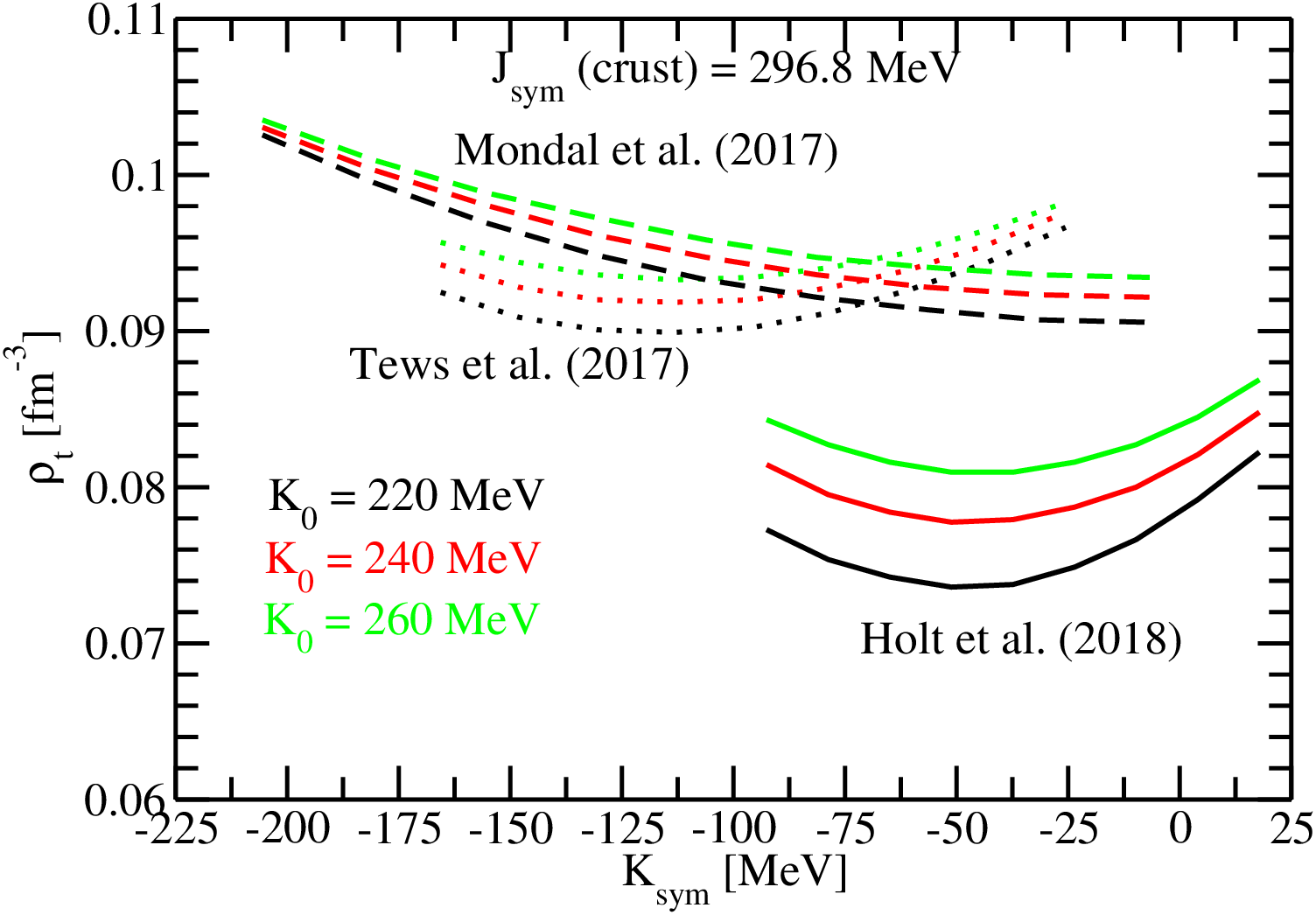}
 }
  \caption{(Color online) The crust-core transition density as a function of $K_{\rm{sym}}$ with the indicated three different $K_{\rm{sym}}-L$ correlations with $K_0=220, 240$ and 260 MeV and 
  $J_{\rm{sym}}(\rm{crust})=-200$ (left) and $+296.8$ MeV (right), respectively.}\label{densityK}
\end{center}
\end{figure*}

Below the crust-core transition density, we use the NV EOS \cite{Negele73} for the inner crust and the BPS EOS \cite{Baym1971} for the outer crust. 
For the purposes of this work, this choice is sufficient. However, we notice that more modern descriptions for both the inner crust including a possible pasta phase, see, e.g. \cite{Newton12}, and the outer crust built from the same
interaction as the core EOS in a uniform approach, see, e.g. \cite{Fortin}, are available in the literature.

Having discussed earlier how to find the crust-core transition point, we now discuss briefly how the core EOS is  determined, namely selecting the values or ranges of the high-density EOS parameters especially the $J_0$ and $J_{\rm sym}$.  The maximum mass 2.14 M$_{\odot}$ of NSs observed so far \cite{Mmax} requires $J_{0}$ to be higher than about $-200$ MeV \cite{Zhang20a} slightly depending on the symmetry energy parameters $L$, $K_{\rm sym}$ and $J_{\rm sym}$ used \cite{Zhang20a}. For the purposes of this work focusing on effects of $K_{\rm sym}-L$ correlation on properties of canonical NSs, it is sufficient to simply use a constant $J_0$ that is large enough to support NSs as massive as about 2.0 M$_{\odot}$ in the whole EOS parameter space considered. Here we present results all obtained with $J_{0}=-180$ MeV. Using different values, e.g., -215 MeV, also satisfying the above conditions and being consistent with the results of Bayesian analyses of both NS properties and heavy-ion reactions mentioned earlier, our results remain qualitatively the same.  

To our best knowledge, presently there is no clear constraint on the value of $J_{\rm sym}$ from neither astrophysical observations not terrestrial experiments. Interestingly, however, the study of Mondal {\it et al.} predicted its most probable value at $J_{\rm sym}=296.8 \pm 73.6$ MeV by putting several terrestrial experimental constraints on the universal correlations of EOS parameters they studied \cite{India17}. For demonstrating effects of the $K_{\rm sym}-L$ correlations, here we adopt the value of $J_{\rm sym}=296.8 $ MeV for both the crust and the core, except for a comparison to examine effects of the crust using a different value of $J_{\rm sym}$. We did systematically vary the values of both $J_0$ and $J_{\rm sym}$ under the conditions that all EOSs have to be casual, dynamically stable and stiff enough to support NSs as massive as about 2.0 M$_{\odot}$, our qualitative conclusions are the same while there are some slight differences quantitatively. Since the crust-core transition density and pressure have some appreciable dependences on $J_{\rm sym}$, to be consistent, we use the same $J_{\rm sym}$ in finding the crust-core transition point and constructing the core EOS unless otherwise specified. 

The resulting EOS for the whole NS in the form of $P(\epsilon)$ is then used as the input in solving the standard Tolman-Oppenheimer-Volkov (TOV) NS structure equations \citep{Tolman1934,Oppenheimer39}
\begin{equation}\label{TOVp}
\frac{dP}{dr}=-\frac{G(m(r)+4\pi r^3P/c^2)(\epsilon+P/c^2)}{r(r-2Gm(r)/c^2)},
\end{equation}
\begin{equation}\label{TOVm}
\frac{dm(r)}{dr}=4\pi\epsilon r^2. 
\end{equation}
The NS mass M is obtained from integrating the mass profile $m(r)$ and the radius R is found when the pressure becomes zero on the surface 
starting from the central pressure $P_c$ where $m(0) = 0$. The codes used in this work are developed from modifying those used in Refs. \cite{Zhang2018,Farooh}.  
\begin{figure*}[htb]
\begin{center}
    \resizebox{0.48\textwidth}{!}{
 \includegraphics[scale=0.4]{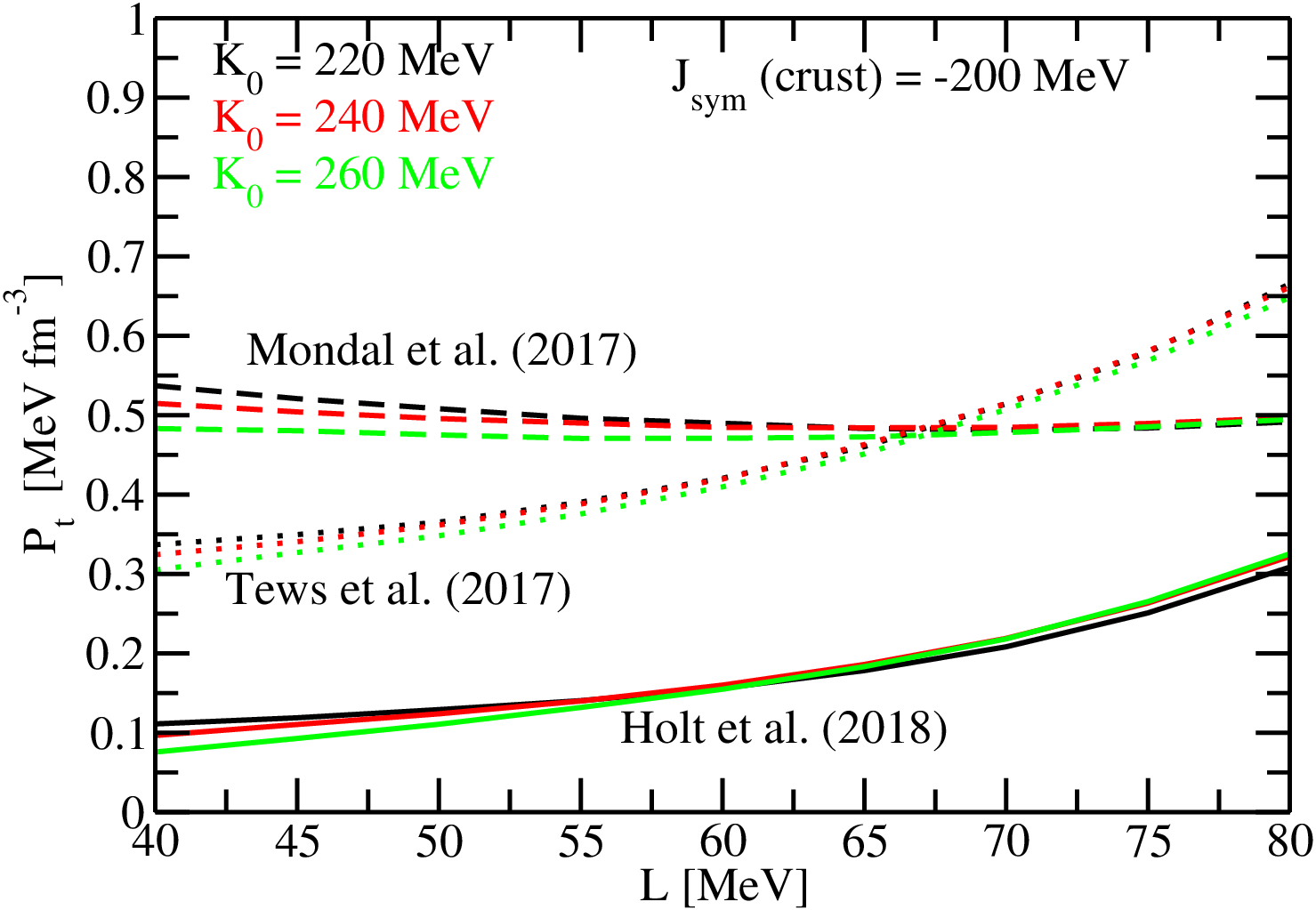}
  }
  \hspace{0.3cm}
     \resizebox{0.48\textwidth}{!}{
  \includegraphics[scale=0.4]{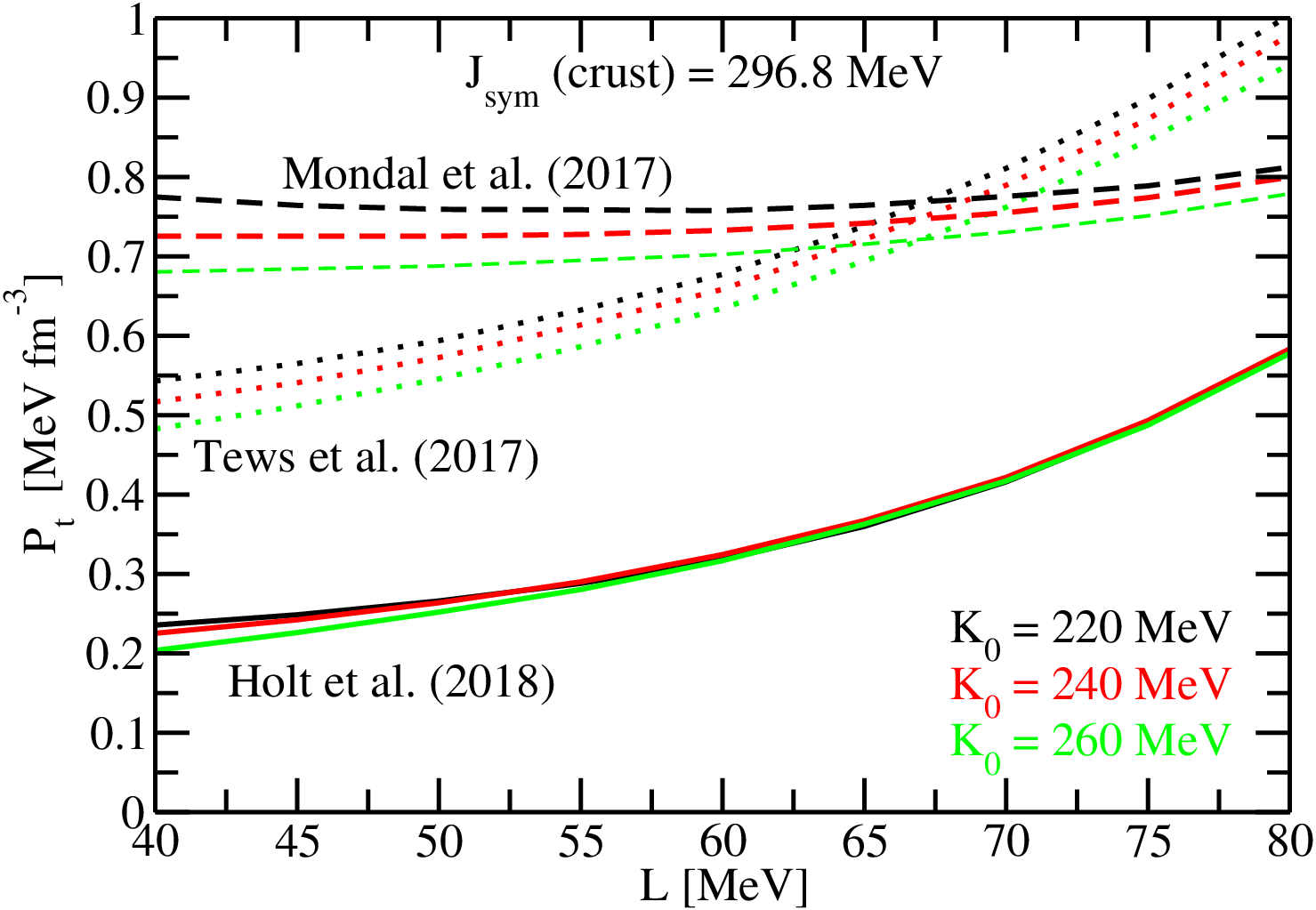}
 }
  \caption{(Color online)The crust-core transition pressure as a function of $L$ with the indicated three different $K_{\rm{sym}}-L$ correlations, $K_0=220, 240$ and 260 MeV, $J_{\rm{sym}}(\rm{crust})=-200$ (left) and $+296.8$ MeV (right), respectively.}\label{PtL}
\end{center}
\end{figure*}
\begin{figure*}[htb]
\begin{center}
\vspace{0.5cm}
    \resizebox{0.48\textwidth}{!}{
 \includegraphics[scale=0.4]{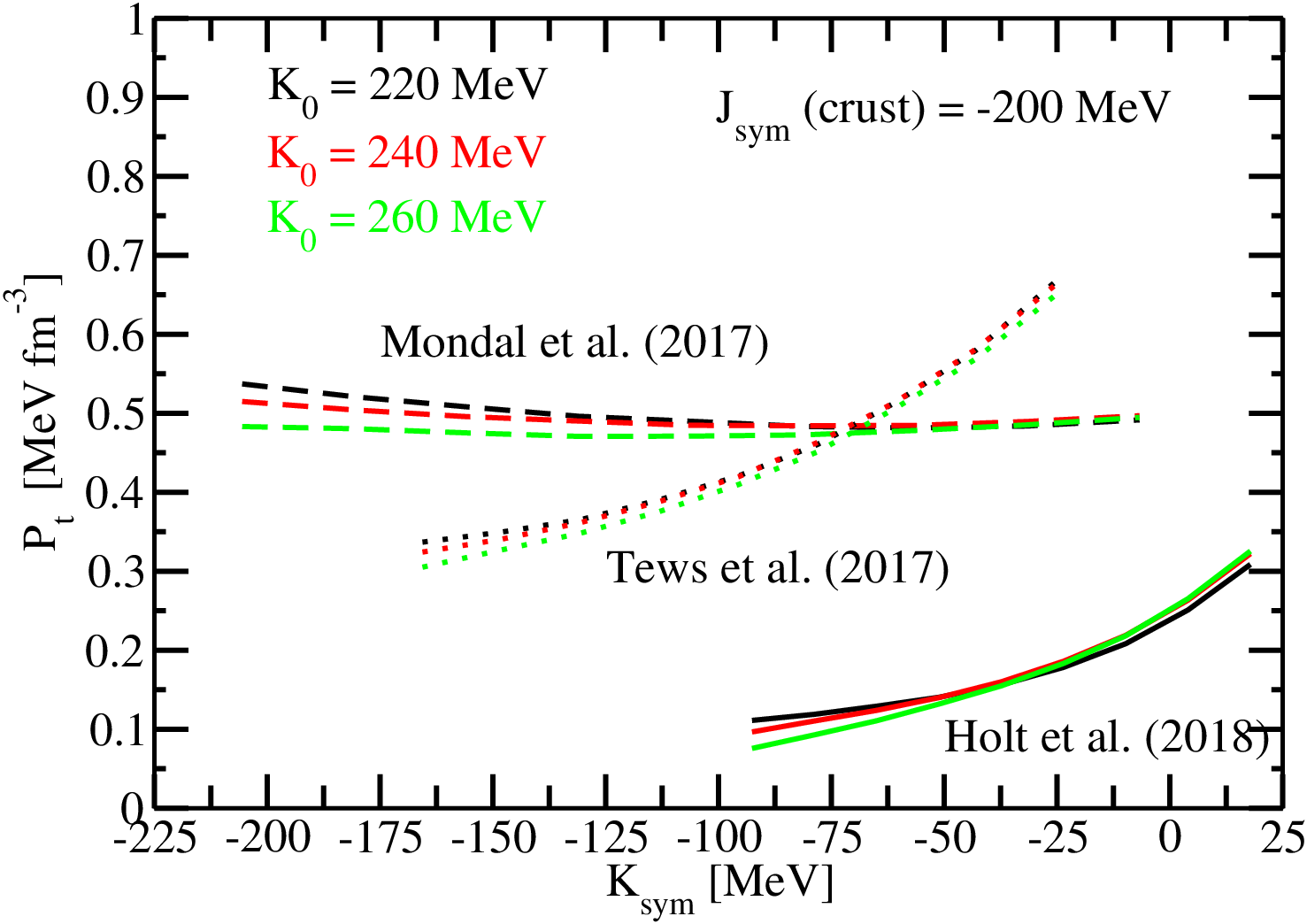}
  }
    \hspace{0.3cm}
     \resizebox{0.48\textwidth}{!}{
  \includegraphics[scale=0.4]{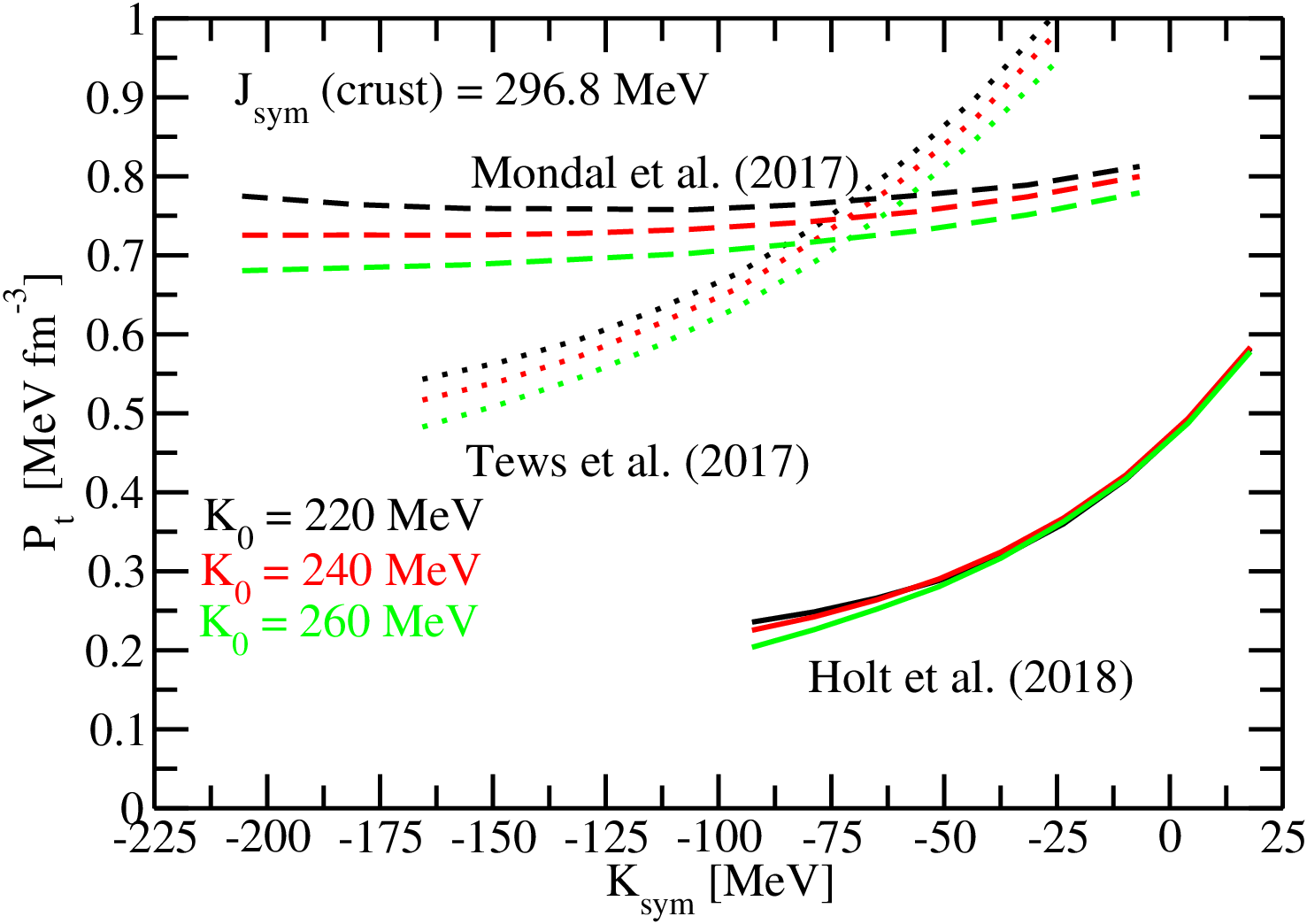}
  }
  \caption{(Color online)The crust-core transition pressure as a function of $K_{\rm{sym}}$ with the indicated three different $K_{\rm{sym}}-L$ correlations, $K_0=220, 240$ and 260 MeV,  $J_{\rm{sym}}(\rm{crust})=-200$ (left) and $+296.8$ MeV (right), respectively.}\label{PtK}
\end{center}
\end{figure*}

\section{Results and discussions}
\subsection{Effects of the $K_{\rm{sym}}-L$ correlation on the NS crust-core transition density}\label{CC}

Shown in Fig. \ref{densityL} are the crust-core transition density as a function of $L$ with the indicated three different $K_{\rm{sym}}-L$ correlations, $K_0=220, 240$ and 260 MeV, $J_{\rm{sym}}(\rm{crust})=-200$ (left) and $+296.8$ MeV (right), respectively. The same results are also shown in Fig. \ref{densityK} but as functions of $K_{\rm{sym}}$ instead. Overall, the crust-core transition density is around $\rho_0/2$ often used as its fiducial value in the literature. 
Clearly, among the variables studied, the $K_{\rm{sym}}-L$ correlation has the strongest effect on the crust-core transition density. 

The incompressibility $K_0$ of symmetric nuclear matter also shows a significant effect especially in the case with the Holt correlation. Its increase makes the transition density higher. This can be understood easily from Eq. (\ref{kmu2}). A higher value of $K_0$ needs a higher $\rho_t$ to make $K_{\mu}$ zero as the first term of Eq. (\ref{kmu2}) involving $K_0$ is always positive and increases with density quadratically. 

It is seen that the $J_{\rm{sym}}$ parameter plays an appreciable role. Its increase also makes the transition density higher through the $E^{-1}_{\rm sym}(\rho)$ term in Eq. (\ref{kmu2}).  As shown in Fig. \ref{cor}, in the same range of L, the three correlations have different ranges for the $K_{\rm{sym}}$ parameter. Since the latter plays the most important role in determining the crust-core transition density, the correlation effects look more obvious in Fig. \ref{densityK} where the transition density is shown as a function of $K_{\rm{sym}}$.  We notice that because the L and $K_{\rm{sym}}$ are correlated, the results shown in Fig. \ref{densityL} and Fig. \ref{densityK} are not 
independent and can be translated into each other easily.

For comparisons, it is worth noting that effects of both the $K_{\rm{sym}}-L$ correlation and $K_0$ examined here are actually larger than those due to the isospin-dependence of the surface tension examined in the compressible liquid drop model \cite{Newton12,Fra-crust2} or the coefficients of the $\delta^4$ and $\delta^6$ terms in expanding the EOS of isospin-asymmetric nuclear matter \cite{JXu1,JXu2,Gon17}. Thus, considering all the factors and their associated uncertainties, it appears that the model-dependent $K_{\rm{sym}}-L$ correlation is a dominating factor in determining the crust-core transition density.
\begin{figure*}[htb]
%\begin{center}
%\vspace{-1cm}
\resizebox{0.75\textwidth}{!}{
 \includegraphics[scale=1]{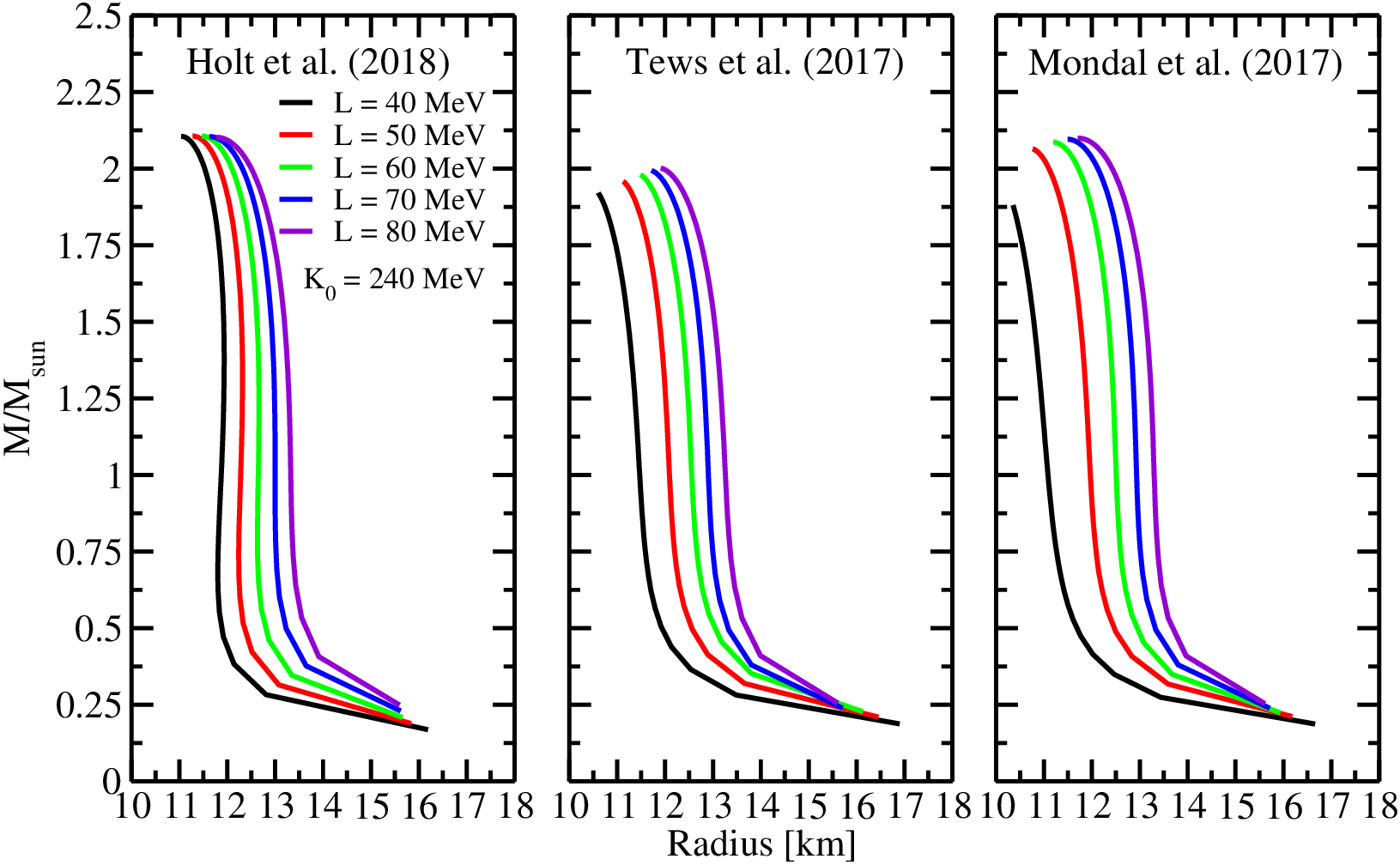}
}
  \caption{(Color online) The mass-radius correlation of neutron stars obtained using the three different $K_{\rm{sym}}-L$ correlations as indicated but the same $K_0=240$ MeV and all other EOS parameters indicated or given in the text.}\label{MR}
\end{figure*}

\subsection{Effects of the $K_{\rm{sym}}-L$ correlation on the crust-core transition pressure}
We notice that in constructing the EOS for the whole NS by connecting the core EOS with that of the crust, the transition density plays the major role while the transition pressure is only used to check if the pressure around the transition point continuously increases with density to ensure the dynamical stability of the NS. Thus, all effects of the different $K_{\rm{sym}}-L$ correlations on the radii and tidal deformations of NSs are coming through the crust-core transition density and the core EOS. Nevertheless, it is interesting to examine how the transition pressure itself depends on the EOS parameters especially the $K_{\rm{sym}}-L$ correlation. 

Moreover, the crust-core transition pressure may play an important role in understanding the still puzzling glitch phenomenon of some pulsars. The crustal fraction of the moment of inertia ${\Delta I}/{I}$ is a quantity that can be extracted from observations of pulsar glitches. It can be expressed approximately in terms of the crust-core transition density and pressure as~\cite{Lattimer2000,Lattimer2001,Lat07}
\begin{eqnarray}\label{dI}
\frac{\Delta I}{I} &\approx& \frac{28\pi P_t R^3}{3 M c^2}
\frac{(1-1.67\xi-0.6\xi^2)}{\xi}\nonumber\\
&\times&\left[1+\frac{2P_t(1+5\xi-14\xi^2)}{\rho_t m c^2 \xi^2}\right]^{-1}
\end{eqnarray}
where $m$ is the baryon mass and $\xi=G M/R c^2$. Analytically, the $K_{\rm{sym}}-L$ correlation may affect significantly the ${\Delta I}/{I}$ through both the transition density and pressure. It is thus also important to examine 
 effects of the $K_{\rm{sym}}-L$ correlation on the crust-core transition pressure.

Having found the crust-core transition density $\rho_t$ and the corresponding isospin asymmetry $\delta_t$ through the charge neutrality and $\beta$-equilibrium conditions \cite{Zhang2018}, one can find the corresponding crust-core transition pressure using the formalism given in Section \ref{theory}. Shown in Fig. \ref{PtL} and Fig. \ref{PtK} are the crust-core transition pressures as functions of $L$ and $K_{\rm{sym}}$, respectively. The other EOS parameters used are the same as those used in calculating the transition density shown in Fig. \ref{densityL} and Fig. \ref{densityK}. It is seen that the trends follow that of the transition density as one expects. However, effects of the $K_0$ are significantly reduced while the strong effects of the $K_{sym}-L$ correlation and the appreciable effects of $J_{\rm{sym}}$ remain. 

In the thermodynamical approach used here, the crust-core transition pressure can be approximated as \cite{Lattimer2000,Lattimer2001,Lat07}
\begin{eqnarray}\label{PT}
   &&P_t \approx \frac{K_0}{9}\frac{\rho_t^2}{\rho_0}\left(\frac{\rho_t}{\rho_0} - 1\right)\\ && + \rho_t \delta_t \left[\frac{1-\delta_t}{2} E_{\rm sym}(\rho_t) + \left(\rho \frac{dE_{\rm sym}(\rho)}{d\rho}\right)_{\rho_t} \delta_t \right].\nonumber
\end{eqnarray}
Besides the explicit dependence on the magnitude and slope of the symmetry energy at $\rho_t$, the latter itself also carries effects of the symmetry energy. Effects of the various EOS parameters are thus intertwined in the transition pressure. Nevertheless, it is clearly seen that the $K_{\rm{sym}}-L$ correlation plays the dominating role in determining the crust-core transition pressure.

\begin{figure*}[htb]
%\vspace{1.cm}
\begin{center}
 \includegraphics[width=0.75\linewidth]{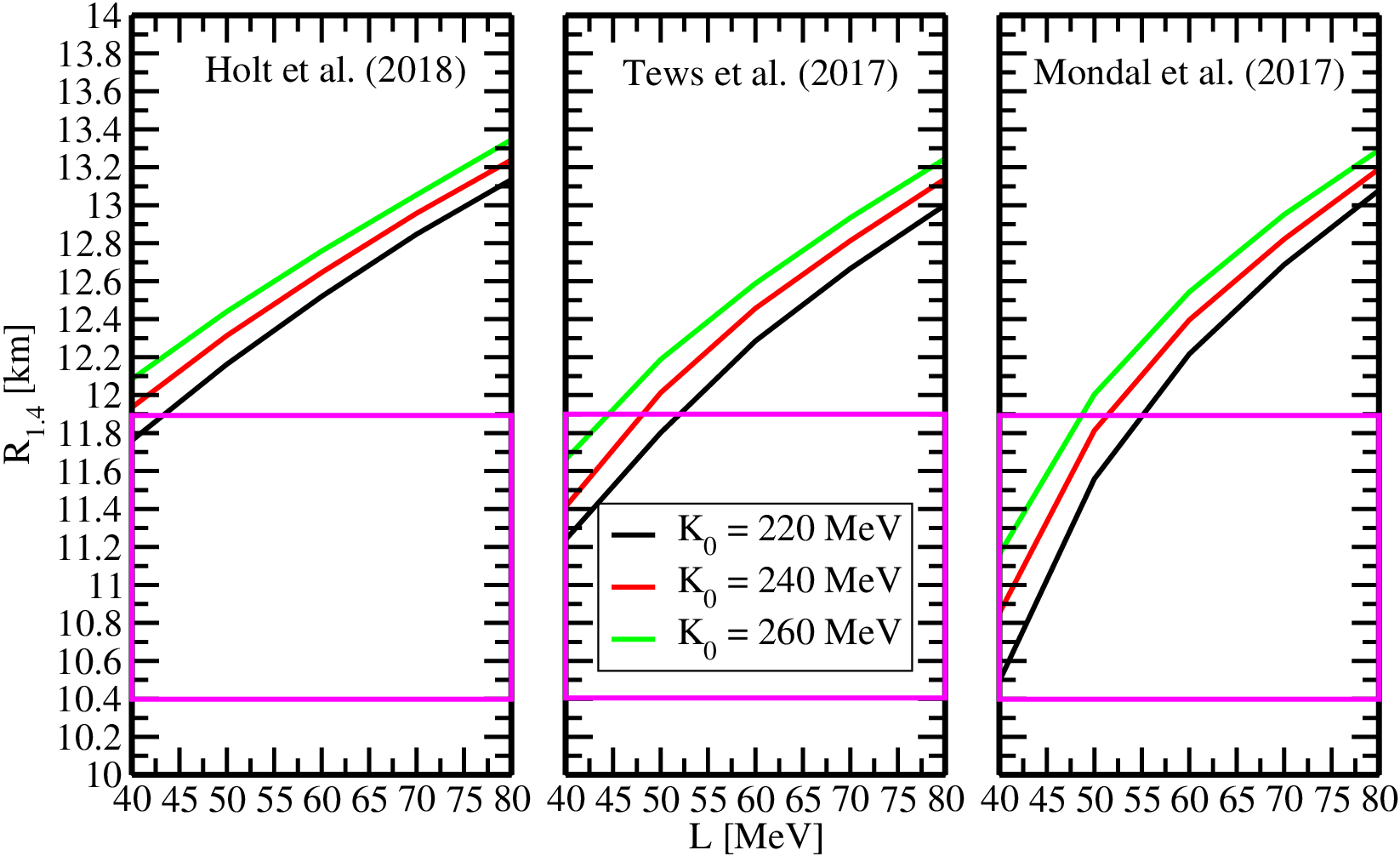}
   \caption{(Color online) The radius $R_{1.4}$ of canonical neutron stars of mass 1.4 M$_{\odot}$ as a function of the symmetry energy slope parameter L obtained using the three different $K_{\rm{sym}}-L$ correlations as indicated and three different values of $K_0$ while keeping all other EOS parameters the same as indicated or given in the text. The magenta box indicate the most probable value of $R_{1.4}=11.0^{+0.9}_{-0.6}$ km at 90\% confidence level from the latest multimessenger observation of GW170817 \cite{Capano20}.}\label{R14}
\end{center}
\end{figure*}

\subsection{Imprints of the $K_{\rm{sym}}-L$ correlation on the radius of canonical neutron stars}\label{MR-r}
We now turn to investigating effects of the $K_{\rm sym}-L$ correlation on some observational properties of canonical NSs of mass 1.4 M$_{\odot}$. We focus on the radii and tidal deformations of these canonical NSs as they are most sensitive to the L and $K_{\rm sym}$ without much influences of the high-order EOS parameters, such as the $J_0$ and $J_{\rm sym}$ \cite{Xie19}. The latter are known to affect significantly the masses and radii of more massive NSs \cite{Zhang2018,Xie20a}. 

Shown in Fig. \ref{MR} are the NS mass-radius (M-R) correlations obtained using the three different $K_{\rm{sym}}-L$ correlations but the same $K_0=240$ MeV and all other EOS parameters discussed earlier. Our results with other values of $K_0$ are similar and the resulting effects on the radius $R_{1.4}$ of canonical neutron stars will be examined later.  Overall, the three correlations lead to generally very similar M-R correlations but there are interesting differences in the L dependence especially at low L values. The increase of $R_{1.4}$ with L is a well known and common feature of all EOSs, see, e.g., Ref. \cite{LiSte}. 

As mentioned earlier, the maximum mass of neutron stars is mostly determined by the SNM EOS characterized by the $K_0$ and $J_0$ parameters. 
While effects of the symmetry energy parameters on the maximum mass are generally small but can be appreciable when the $K_0$ and $J_0$ are fixed. Comparing the results in the three windows with different $K_{\rm{sym}}-L$ correlations, it is seen that both the maximum mass and the radius $R_{1.4}$ are appreciably different at low L values. This can be well understood from the $K_{\rm{sym}}-L$ correlations shown in Fig. \ref{cor}. For low L values, the $K_{\rm{sym}}$ values are the lowest for the Mondal correlation but the highest for the Holt case. The latter thus has the stiffest while the Modal case has the softest symmetry energy, making the strongest and the weakest contribution to the nuclear pressure, respectively. Consequently, for low L values, the Holt correlation predicts a higher value for the maximum mass and also a larger radius compared to the other two cases. While it is the opposite for the Mondal case. 
As we discussed in Section \ref{kmu0}, in the Mondal and Tew cases, the second dynamical instability may happen at high densities when the resulting symmetry energy is super-soft with small L values but big negative $K_{\rm sym}$ values especially if $K_0$ is also small. Without introducing new phases to stabilize neutron star matter above the onset density of the second dynamical instability, the maximum mass supported by the Mondal and Tews correlations are thus smaller compared to the Holt result in cases where both the L and $K_0$ values are small towards their lower limits currently known. 

The effects of different $K_{\rm{sym}}-L$ correlations on the radius $R_{1.4}$ of canonical NSs can be seen more clearly in Fig. \ref{R14} where the $R_{1.4}$ is shown as a function of L with three different $K_0$ values covering their current uncertainty ranges.  As mentioned earlier, the high-density EOS parameters $J_0$ and $J_{{\rm sym}}$ have little effects on the $R_{1.4}$ \cite{Zhang2018,Zhang19apj,Zhang19EPJA,Xie19,Xie20a}. We thus focus on the dependence of $R_{1.4}$ on L and $K_0$ using the three different $K_{\rm{sym}}-L$ correlations. A recent study combining multimessenger observations of GW170817 and many-body theory predictions using nuclear forces based on the chiral effective field theory found that the most probable $R_{1.4}$ is $R_{1.4}=11.0^{+0.9}_{-0.6}$ km at 90\% confidence level \cite{Capano20}. The latter is indicated with the magenta boxes in  Fig. \ref{R14} for comparisons.

Several interesting observations can be made from comparing the results in the three windows:
\begin{itemize}
\item
The $R_{1.4}$ increases almost linearly with both L and $K_0$ in their respective uncertainty ranges. The dual dependence of $R_{1.4}$ on L and $K_0$ indicates that it is important to have the prior knowledge of $K_0$ as accurately as possible to infer precisely the value of L from NS radius measurements. On the other hand, it has been known for a long time that the remaining uncertainty of about $\pm 20$ MeV in extracting $K_0$ from studying giant resonances of finite nuclei is mainly due to the correlations of $K_0$ and parameters characterizing the $E_{\rm sym}(\rho)$ near $\rho_0$, see, e.g., Refs. \cite{Colo08,MM3} for detailed discussions. It is thus interesting to note that Bayesian inferences of multiple EOS parameters simultaneously from combined data of astrophysical observations and nuclear experiments have the potential to further pin down the L and $K_0$ parameters all together \cite{Xie19,Xie20a}.

\item
For all three $K_{\rm{sym}}-L$ correlations considered, the $R_{1.4}$ is about the same when L is higher than about 60 MeV.
This is because at higher L values, the three $K_{\rm{sym}}-L$ correlations largely overlap as shown in Fig. \ref{cor}. The $R_{1.4}$ becomes gradually more different as the L decreases. The Holt correlation has the stiffest symmetry energy leading to the largest radius while the Mondal case has the softest symmetry energy giving the smallest $R_{1.4}$ value as we discussed earlier. Moreover, effects of $K_0$ on $R_{1.4}$ is almost independent of the 
$K_{\rm{sym}}-L$ correlation. At a given L value, the $R_{1.4}$ increases with increasing $K_0$ as one expects. Quantitatively, however, the uncertainty of $R_{1.4}$ due to that of $K_0$ is only about 4\% while that due to the 
uncertainty of L is about 16\%. For a comparison, at L=40 MeV with $K_0=220$ MeV, the difference in $R_{1.4}$ from using the Holt and Mondal $K_{\rm{sym}}-L$ correlation is about 12\%. This is actually slightly less than the approximately 14\% uncertainty of the latest constrain on $R_{1.4}$ from the multimessengers observations of GW170817. Thus, the latter can not distinguish the three $K_{\rm{sym}}-L$ correlations.

\item
The observational constraint $R_{1.4}=11.0^{+0.9}_{-0.6}$ km can put some useful limits on the slope L. The Holt correlation requires the L parameter to be less than about 45 MeV close to the lower boundary of its fiducial value. While the other two $K_{\rm{sym}}-L$ correlations prefers an upper limit of L in the range of 45-55 MeV for $K_0$ between 260 and 220 MeV. This information is useful for more accurate extraction of EOS parameters and understanding their model dependences in future analyses of more data to be available.
\end{itemize}

\begin{figure}[htb]
%\vspace{0.5cm}
\begin{center}
 \includegraphics[width=1\linewidth]{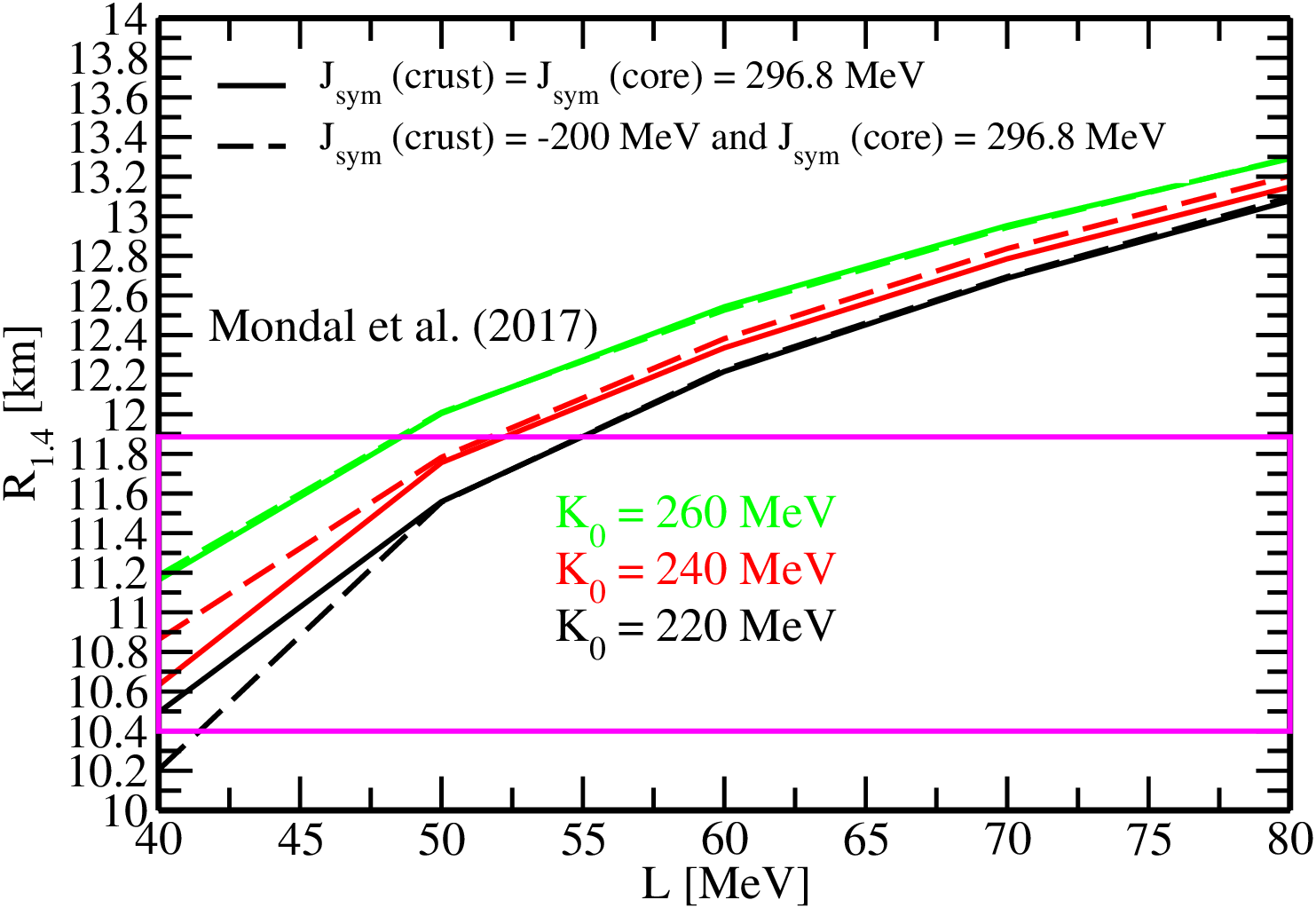}
  \caption{(Color online) Solid: The default calculation (with the same $J_{\rm{sym}}$ of 296.8 MeV for the core EOS and for finding the crust-core transition point). Dashed: a calculation using 
  $J_{\rm{sym}}=-200$ MeV for finding the crust-core transition point but still the default $J_{\rm{sym}}$ for the core EOS with the Mondal correlation.}\label{comp}
\end{center}
\end{figure}

The $K_{\rm{sym}}-L$ correlation affects both the crust-core transition point and the core EOS. We have shown earlier how the crust-core transition density and pressure are also being affected appreciably by 
the uncertain $J_{\rm{sym}}$ parameter characterizing the $E_{\rm sym}(\rho)$ far away from $\rho_0$ at either sub-saturation or supra-saturation densities \cite{Xie20a}. More quantatively, from $J_{\rm{sym}}=-200$ MeV to $+296.8$ MeV, both the transition density and pressure have appreciable changes of about 10\% when the L is small but there is not much difference when L is large.

To understand effects of $J_{\rm{sym}}$ on the $R_{1.4}$ we have done systematical studies by using different combinations of $J_{\rm{sym}}$ parameters in calculating the crust-core transition point and the core EOS. We found that the effects are negligibly small except when the L value is small close to the lower boundary of its fiducial value. As an example, shown in Fig. \ref{comp} is a comparison of the default calculation (with the same $J_{\rm{sym}}$ of 296.8 MeV for the core EOS and for finding the crust-core transition point) and a calculation that uses $J_{\rm{sym}}=-200$ MeV for finding the 
 crust-core transition point (while the core EOS still uses the default $J_{\rm{sym}}$ of 296.8 MeV) with the Mondal correlation. It is seen that the $R_{1.4}$ versus L results from the two calculation are not much different except when the L  becomes smaller than about 50 MeV and $K_0$ is also small. When the L becomes small, its contribution to the pressure is smaller. Then, what values one use for the high-order symmetry energy parameter $J_{\rm{sym}}$ in finding where to connect the core EOS to the crust EOS becomes more important. While this is the same region of L where the $K_{\rm{sym}}-L$ correlation plays a significant role in determining the $R_{1.4}$, obviously the $K_{\rm{sym}}-L$ correlation effect is much stronger than that due to uncertainty of the $J_{\rm{sym}}$ parameter. Moreover, the comparison indicates that the $K_{\rm{sym}}-L$ correlation effects on the $R_{1.4}$ in the default calculations shown in Figs. \ref{MR} and \ref{R14} come mainly through the core EOS. 

\begin{figure*}[htb]
%\begin{center}
%\vspace{1cm}
\resizebox{0.7\textwidth}{!}{
 \includegraphics[scale=0.9]{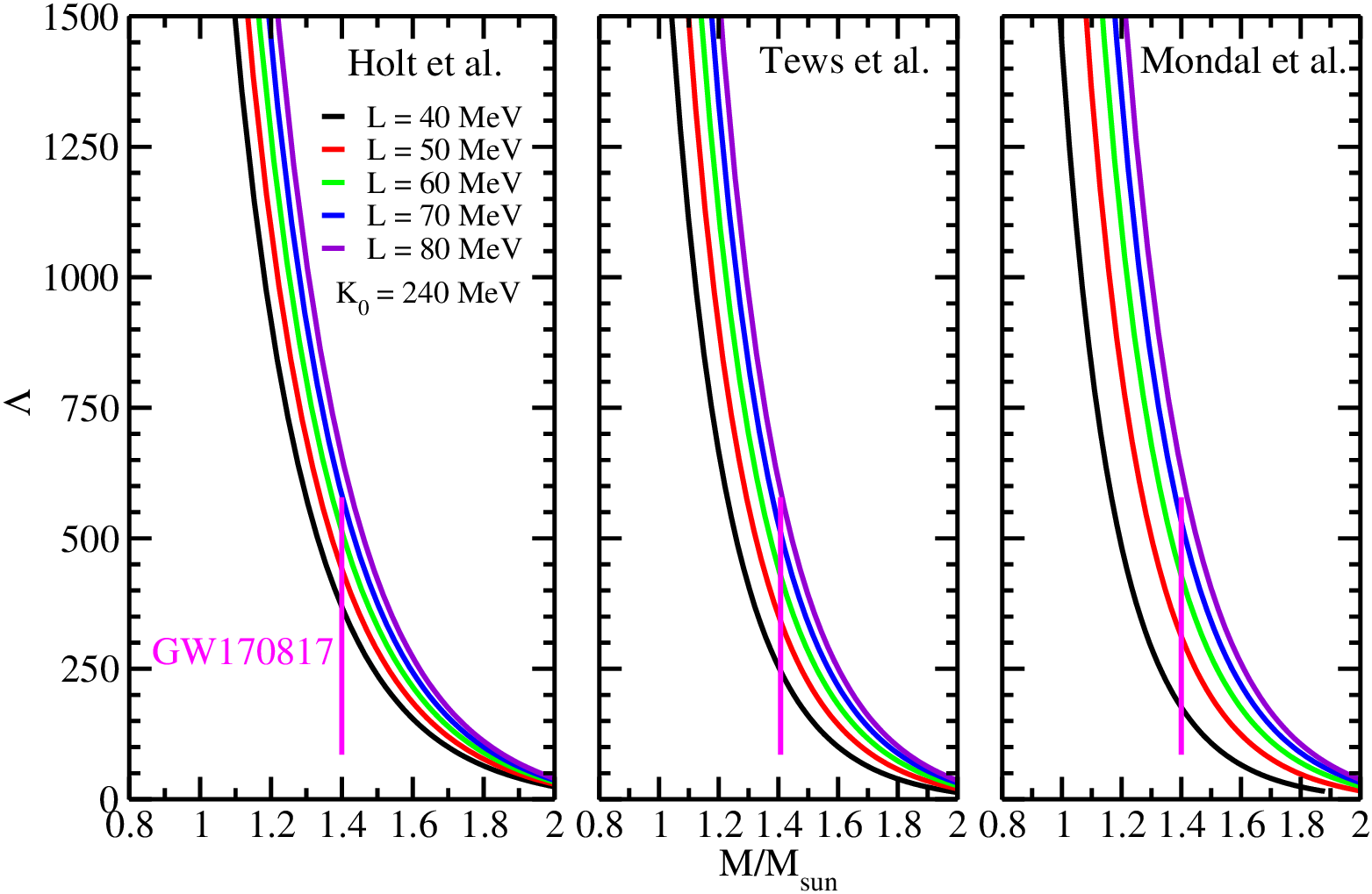}
}
  \caption{(Color online) The scaled tidal deformability as a function of mass of neutron stars obtained using the three different $K_{\rm{sym}}-L$ correlations as indicated but the same $K_0=240$ MeV and all other EOS parameters indicated or given in the text. The magenta bar between 70 and 580 is the tidal deformability of canonical neutron stars extracted from LIGO/VIRGO's observation of GW170817 \cite{LIGO}.}\label{LM}
%\end{center}
\end{figure*}
\begin{figure*}[htb]
\begin{center}
\vspace{0.85cm}
 \resizebox{0.7\textwidth}{!}{
 \includegraphics[scale=0.8]{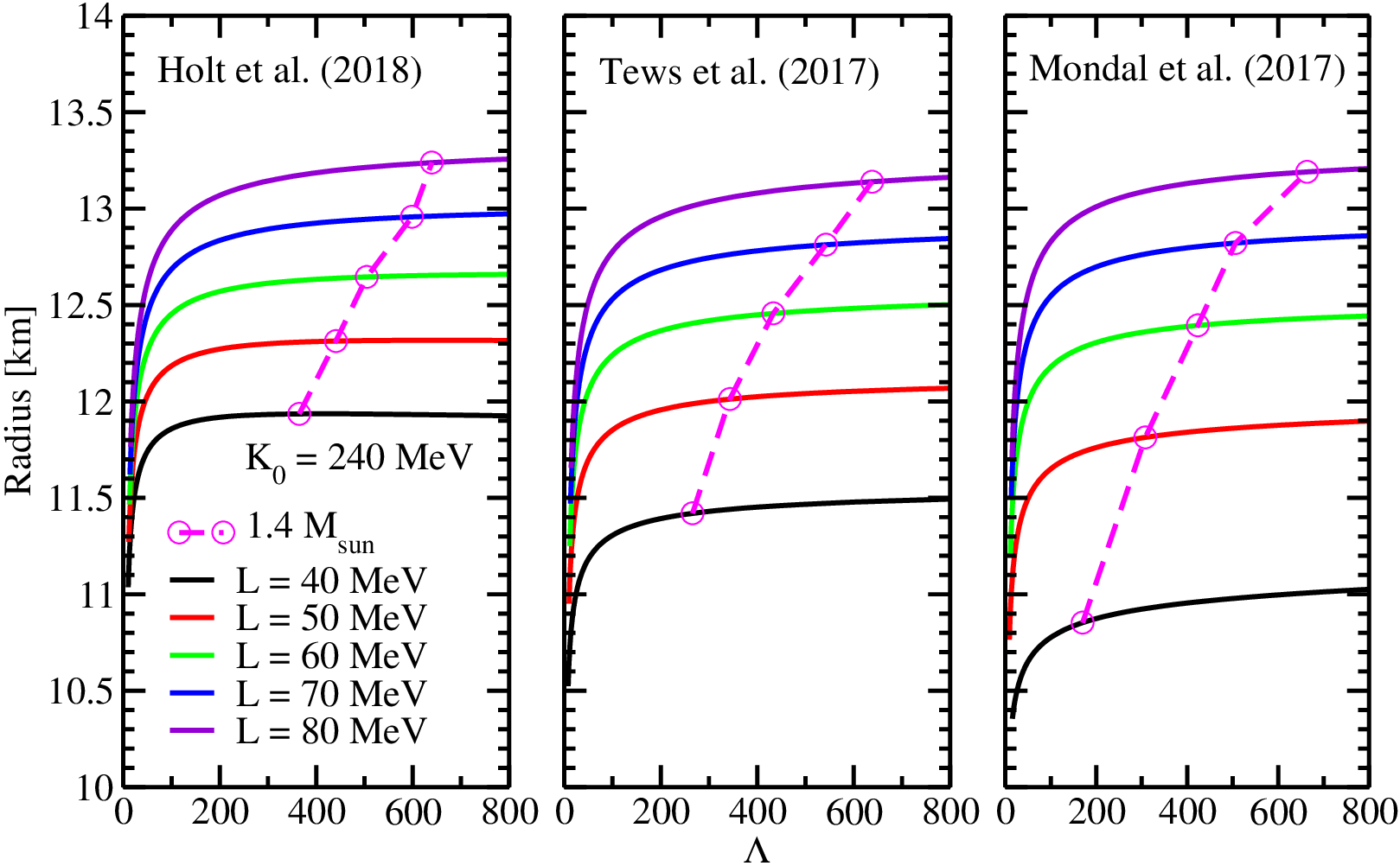}
 }
  \caption{(Color online) The radius-tidal deformability correlations of neutron stars with all masses obtained using the three different $K_{\rm{sym}}-L$ correlations as indicated but the same $K_0=240$ MeV. The magenta circles linked on the dashed lines are for canonical neutron stars.}\label{RL}
\end{center}
\end{figure*}

\subsection{Imprints of the $K_{\rm{sym}}-L$ correlation on the tidal deformation and its correlation with the radius of canonical neutron stars}
We now examine imprints of the $K_{\rm{sym}}-L$ correlation on neutron stars' scaled tidal deformability 
\begin{equation}
\Lambda=\frac{2}{3}k_2 [(c^2/G)R/M]^5
\end{equation}
where $k_2$ is the second Love number obtained from solving coupled differential equations simultaneously with the TOV equation \cite{Flanagan2008,Damour2009,Damour2010,Hinderer2010}. Shown in Fig. \ref{LM} is $\Lambda$ as a function of mass obtained using the three different $K_{\rm{sym}}-L$ correlations but all the same EOS parameters. The magenta bar between 70 and 580 is the tidal deformability of canonical neutron stars from LIGO/VIRGO's observation of GW170817 \cite{LIGO}. Similar to the mass-radius correlation shown in Fig.\ \ref{MR}, the $K_{\rm{sym}}-L$ 
correlation has some observable influences on the tidal deformability for small L values. Again, it can be easily explained by the different $K_{\rm{sym}}$ values with the different $K_{\rm{sym}}-L$ correlation when L is small, as shown in Fig. \ref{cor}. Unfortunately, the range of $\Lambda$ extracted by LIGO/VIRGO from GW170817 is still too big to set a firmer limit on L than its fiducial range. It also can not differentiate the different $K_{\rm{sym}}-L$ correlations. Namely, the use of different $K_{\rm{sym}}-L$ correlations does not affect what one extracts about the symmetry energy from the observational constraint on $\Lambda$ alone from GW170817.

Effects of the $K_{\rm{sym}}-L$ correlation can be more clearly seen in the $R-\Lambda$ correlation by combing the $M-R$ plot of Fig.\ \ref{MR} and the $\Lambda-M$ plot of Fig. \ref{LM}. In the $R-\Lambda$ correlation plot shown in Fig. \ref{RL}, each point on a given cure with a fixed L has a specific mass. Starting from the first point on the left, the mass decreases continuously on any curve with the same L. For the very massive neutron stars with masses around 2.0 to 1.8 M$_{\odot}$ on each curve the radius increases monotonically with increasing $\Lambda$ (decreasing mass) until the plateau is reached. On the plateau, namely in a large range of mass around 1.4 M$_{\odot}$ from approximately 1.8 down to 0.5 M$_{\odot}$, the radius stays approximately a constant while the $\Lambda$ keeps increasing. 

The magenta circles linked on the dashed lines are for canonical neutron stars of mass 1.4 M$_{\odot}$ on curves with different L values. It is seen that for these canonical NSs of the same mass, the radius increases with $\Lambda$ approximately linearly. It is known that the upper limit of $\Lambda$ for canonical NSs from observing GW170817 is more reliable than its lower limit which is more model dependent \cite{LIGO}. While the two NSs involved in GW170817 have significantly different mass ranges, two independent analyses \citep{LIGO,De18} using different approaches all found consistently that the two NSs have essentially identical radii. It is thus appropriate for the discussions here to assume the NS mass is 1.4 M$_{\odot}$. Comparing results in the three windows and using the $\Lambda_{\rm maximum}=580$, it is seen that all three  $K_{\rm{sym}}-L$ correlations give consistently the same upper limit of $R_{1.4}\leq 13.2$ Km and the corresponding upper limit of $L\leq 80$ MeV. This is consistent with the results of recent Bayesian analyses \cite{Xie19,Xie20a} of the LIGO/VIRGO data as we mentioned in the introduction. Again, as the L decreases, effects of the  $K_{\rm{sym}}-L$ correlation becomes more obvious. Unfortunately, the lower limit of $\Lambda$ from GW170817 is currently unreliable. In fact, the lower limit of $\Lambda_{\rm minimum}=70$ requires L values much smaller than the lower boundary of its fiducial value discussed in the introduction. 

In principle, independent measurements of $R_{1.4}$ and $\Lambda_{1.4}$ for canonical NSs will put a more stringent constraint on L and possibly also on the $K_{\rm{sym}}-L$ correlation. While many recent works in the literature have focused on extracting the $R_{1.4}$ from $\Lambda_{1.4}$ using the GW170817 data, there are independent measurements of $R_{1.4}$ from other observations, such as X-ray observations. 
Since the constraining box on $R_{1.4}$ shown in Fig. \ref{RL} already used the $\Lambda$ from GW170817 as one of the multimessengers, lets examine how the NICER's recent measurement of PSR J0030+0451 using X-rays may help. The NICER Collaboration measured simultaneously both the mass and radius of PSR J0030+0451. Their results from two somewhat independent analyses are: $M=1.44^{+0.15}_{-0.14}$ M$_{\odot}$ and $R=13.02^{+1.24}_{-1.06}$ km \citep{Miller19}, and $M=1.34^{+0.16}_{-0.15}$ M$_{\odot}$ and $R=12.71^{+1.19}_{-1.14}$ km \citep{Riley19} at 68\% confidence level. For a qualitative discussion, one can safely assume the mass is about 1.4 M$_{\odot}$. The most probable radii from both analyses are consistent with the upper radius indicated by the upper limit of $\Lambda_{1.4}$ from LIGO/VIRGO as we discussed above. However, NICER's 68\% upper radius boundary is as high as 14.26 km or 13.9 km, allowing much higher L values than that allowed by the LIGO/VIRGO data beyond the limit of its fiducial value. Nevertheless, NICER's lower radius limit $R_{1.4} (\rm minimum)$ from the two analyses, i.e., $R_{1.4} (\rm minimum)$=11.96 km or 11.57 km, can put a useful lower limit $L_{\rm minimum}$ on L. This limit, however, depends on the $K_{\rm{sym}}-L$ correlation one uses. It is seen from Fig. \ref{RL} that $L_{\rm minimum}$ is between 40-50 MeV, 50-60 MeV and 50-60 MeV for the Holt, Tews and Mondal correlations, respectively. As we discussed before, the last two have  approximately the same $K_{\rm{sym}}-L$ correlation from the same sets of model predictions. Thus, the different $K_{\rm{sym}}-L$ correlations considered affect the extraction of $L_{\rm minimum}$ by above 10 MeV. 

Overall, the most probable value of $R_{1.4}$ from NICER can independently limit the most probable value of L to the range of 40-80 MeV consistent with its known fiducial range. This is also consistent with the finding of the detailed Bayesian analyses of the combined LIGO/VIRGO and NICER data \cite{Xie20a}. Certainly, more coming independent data for both the $R_{1.4}$ and $\Lambda_{1.4}$ will help further constrain the value of L and the $K_{\rm{sym}}-L$ correlation. 

\section{Summary and conclusions}
In summary, using a meta-model of nuclear EOSs we examined effects of nuclear EOS parameters especially the curvature ($K_{\rm{sym}}$)-slope (L) correlation of nuclear symmetry energy on the crust-core transition density and pressure in neutron stars. We also examined imprints of the $K_{\rm{sym}}-L$ correlation on astrophysical observables especially the radius and tidal deformability of canonical neutron stars. \\

Our main conclusions are the following:
\begin{itemize}
\item The crust-core transition density and pressure have some appreciable dependences on the incompressibility $K_0$ but are insensitive to the skewness $J_0$ of symmetric nuclear matter.
\item The crust-core transition density and pressure are sensitive to the $L$, $K_{sym}$ and $J_{sym}$ parameters independently as well as the $K_{sym}-L$ correlation.
\item The curvature $K_{sym}$ plays a more important role than the slope $L$ in determining the crust-core transition density.
\item The $J_0$ and $J_{sym}$ parameters have little effects on $R_{1.4}$ and the $K_{sym} - L$ correlation effects come through the core EOS.
\item The $K_{\rm{sym}}-L$ correlation has strong imprints on the radius and tidal deformability of canonical neutron stars especially when the slope L is close to the lower limit (40 MeV) of its currently known fiducial value. 
\end{itemize}

The astrophysical imprints of $K_{\rm{sym}}-L$ correlation can potentially help better constrain the poorly known high-density behavior of nuclear symmetry energy.  
In particular, if a unique $K_{\rm{sym}}-L$ correlation can be firmly established by observations/experiments, it will facilitate the extraction of the very poorly known $K_{\rm{sym}}$ parameter progressively from the relatively better determined L value. We thus also examined whether existing tidal deformability data from LIGO/VIRGO's observation of GW170817 an the NS radius data from NICER's recent observation of PSR J0030+0451 can help distinguish the three different $K_{\rm{sym}}-L$ correlations considered, and how well they can constrain the L parameter. Consistently with earlier findings in the literature, they can put some useful constraints on L. Unfortunately, they can not distinguish the three different $K_{\rm{sym}}-L$ correlations studied. Nevertheless, more precise measurements of especially independent radius and tidal deformability data from multiple observables holds the strong promise of pinning down the curvature-slope correlation, thus help constrain the high-density behavior of nuclear symmetry energy.\\

\noindent{\bf Acknowledgments}\\
We thank Drs. Wen-Jie Xie and Nai-Bo Zhang for helpful discussions. This work was supported in part by the U.S. Department of Energy, Office of Science, under Award Number DE-SC0013702, the CUSTIPEN (China-U.S. Theory Institute for Physics with Exotic Nuclei) under the US Department of Energy Grant No. DE-SC0009971, and a NASA Texas Space Grant Consortium Graduate Fellowship.
%\newpage

\end{document}